\documentclass[12pt,fleqn]{article}

\usepackage{amsmath, amssymb}
\usepackage{graphicx}
\usepackage{epic,eepic}

\newtheorem{theorem}{Theorem}

\newcommand* {\manifield}{\,\raisebox{0.2ex}
 {$\!\stackrel{\textmd{\tiny{(SL)}}}{\Psi}$\!\! }}
\newcommand* {\conjmanifield}{\!\stackrel{\rm
  \raisebox{-0.0ex}{\tiny (SL)}}{\ol \Psi}\!\!}
\newcommand* \ket[1]{{\vert#1 \rangle}}

\newcommand* \bra[1]{{\langle #1\vert}}
\newcommand* \bigket[1]{{\left\vert  \begin{matrix} #1
                       \end{matrix} \right\rangle}}
\newcommand* \bigbra[1]{{\left\langle \begin{matrix} #1
                       \end{matrix} \right\vert}}
\newcommand \fract[2]{\textmd{\scriptsize{$\frac{#1}{#2}$}}}

\renewcommand{\a}{\alpha}
\renewcommand{\b}{\beta}
  \newcommand{\g}{\gamma}
\renewcommand{\d}{\delta}
  
\renewcommand{\k}{\kappa}
\renewcommand{\l}{\lambda}
  \newcommand{\m}{\mu}
  \newcommand{\n}{\nu}
  \newcommand{\ve}{\varepsilon}
  \newcommand{\ol}{\overline}

  \newcommand{\G}{\Gamma}
  \newcommand{\RR}{\mathbb{R}}

\begin{document}

\title{From Poincar\'e to affine invariance: How does the Dirac
equation generalize?}

\author{Ingo Kirsch\thanks{Present address:
Humboldt-Universit\"at zu Berlin, Institut f\"ur Physik, D-10115 Berlin,
Germany, email: ik@physik.hu-berlin.de}\\Inst.\ Theor.\
 Physics, University of Cologne\\50923 K\"oln, Germany\\ \\ Djordje\
 $\rm{\check{S}ija\check{c}ki}$\thanks{email: sijacki@phy.bg.ac.yu}\\
 Institute of Physics, P.O.Box 57\\ 11001 Belgrade, Yugoslavia }

\date{}
\maketitle

\begin{abstract} {A generalization of the Dirac equation to the case of
    affine symmetry, with $\ol{SL}(4,\RR)$ replacing $\ol{SO}(1,3)$, is
    considered. A detailed analysis of a Dirac-type Poincar\'e-covariant
    equation for any spin $j$ is carried out, and the related general
    interlocking scheme fulfilling all physical requirements is
    established.
    Embedding of the corresponding Lorentz fields into infinite-component
    $\ol{SL}(4,\RR)$ fermionic fields, the constraints on the
    $\ol{SL}(4,\RR)$
    vector-operator generalizing Dirac's $\gamma$ matrices, as well as the
    minimal coupling to (Metric-)Affine gravity are studied. Finally, a
    symmetry breaking scenario for $\ol{SA}(4,\RR)$ is presented which
    preserves the Poincar\'e symmetry.}
\end{abstract}

\newpage
\section{Introduction}
The outstanding success of the Dirac equation is unprecedented.  It is a
Poincar\'e invariant linear field equation which describes relativistic
spin $\frac{1}{2}$ particles. Interactions can naturally be
introduced by the minimal coupling prescription. In particular, already
in the early stage of its applications, the coupling to the
electro-magnetic field led to many experimental verifications. Nowadays,
it represents one of the key-stones of the Standard model of Electro-Weak
and Strong interactions of elementary particles. In this paper we go
however beyond Poincar\'e invariance and study {\em affine} invariant
generalizations of the Dirac equation, i.e., in other words, a
generalization that will describe a spinorial field in a generic curved
spacetime $(L_4, g)$, characterized by arbitrary torsion and
general-linear curvature. Note that the spinorial fields in the
non-affine generalizations of GR (which are based on higher-dimensional
orthogonal-type generalizations of the Lorentz group) are only allowed
for special spacetime configurations and fail to extend to the generic
case.

As General Relativity is set upon the principle of general covariance,
its fundamental group is the group of diffeomorphisms $Diff(4,\RR)$. A
general-relativization of the concept of spin requires (double-valued)
spinorial representations of $Diff(4,\RR)$, i.e.\ one is interested in
single-valued representations of the double-covering $\ol{Diff}(4,\RR)$.
For a long time it had been wrongly believed that only single-valued
representations of the Lorentz group, vectors and tensors, have a natural
extension to the group $GL(n,\RR)$. However, in 1977 Y.\ Ne'eman has
pointed out \cite{Neem78} that a double-covering $\ol{GL}(n,\RR)$ does
exist, for proof see Ref.\ \cite{Sijacki87}.  The latter in turn contains
spinor representations. The groups $\ol{SL}(n,\RR) \subset
\ol{GL}(n,\RR)$, $n \geq 3$, are necessarily defined in infinite
dimensional vector spaces. Their representations induce those of
$\ol{Diff}(n,\RR)$ \cite{Sijacki87}.

In contradistinction to the tensorial case where one utilizes linear
representations of the group $GL(4,\RR)$ $\subset$  $Diff(4,\RR)$ 
both in the flat tangent (Special Relativity) and in the curved spacetime
(General Relativity), there are two customary constructions that provide
ways to define finite spinors in a curved spacetime \cite{Sija98}: i) One
introduces a set of anholonomic tetrads and defines an action of the
(local) Lorentz group in the tangent space, or ii) One makes use of
nonlinear realizations of the $\overline{Diff}(4,\RR)$ group which are
linear when restricted to the Lorentz subgroup. In both cases spinorial
fields essentially ``live'' in the tangent spacetime.

This asymmetry of treating tensors and spinors in GR is somewhat
unsatisfying from a mathematical point of view. A unified description of
both tensorial and spinorial fields can only be achieved by enlarging the
tangential Lorentz group to the whole linear group which, together with
translations, forms the affine group. The metric-affine gauge theory of
gravity \cite{MAG} appears to be the natural framework for this
unification.

Moreover, the very existence of the $\ol{Diff}(4,\RR)$ fundamental
fermionic fields opens up new roads to studies of the Gravitational
interactions of the fermionic matter at the quantum level (e.g.\ falling
of the proton into a Black Hole, when the thorough recurrences of the
proton Regge trajectory can be excited gravitationally and play an
essential role).

Affine-invariant extensions of the Dirac equation have been studied in
\cite{Mick, Cant, MAG, Neem}. Mickelsson \cite{Mick} has constructed a
$\ol{GL}(4,\RR)$ covariant extension of the Dirac equation. However, its
physical interpretation is rather unclear - in particular, the physically
essential questions: i) the $\ol{GL}(4,\RR)$ irreducible representations
content of the overall representation space and its unitarity features,
and ii) the representation content of the $\ol{SO}(1,3)$ $\supset$
$\ol{SO}(3)$ and/or $\ol{SO}(1,3)$ $\supset$ $\ol{E}(2)$ subgroup-chains
that define the physical particle states were not addressed at all. Cant
and Ne'eman \cite{Cant} found a Dirac-type equation for manifields
(infinite-component fields of $\ol{SL}(4,\RR)$) which is still Poincar\'e
invariant. They use only a subclass of representations of
$\ol{SL}(4,\RR)$, the multiplicity-free ones. Since this class does not
allow a $\ol{SL}(4,\RR)$ vector operator, their field equation cannot be
extended to an affine Dirac-type wave equation.

We will not derive an affine Dirac-type equation explicitly. In this
paper we merely focus on its reduction under the Lorentz group, i.e.\ on
its appearance after the symmetry breaking down to $\ol{SO}(1,3)$, as
well as to the relevant requirements yielding a physically feasible
theory. Nonetheless, in section\mbox{ 2}, we review some general
requirements of a $\ol{SL}(4,\RR)$ vector operator which generalizes
Dirac's $\gamma$-matrices. In this context we find that the mass term in
an affine equation must vanish.  In sections 3 to 5, we investigate
Poincar\'e invariant Dirac-type equations for particles with arbitrary
half-integral spin. We show how the method of Gel'fand et al.\
\cite{Gelf} can be generalized to derive $\gamma$-matrices for these
equations. We state a theorem which yields the minimal sets of
irreducible Lorentz representations needed in such equations.

In section 6, we start our construction of a Poincar\'e invariant
Dirac-type wave equation for manifields. This will be an equation of
the form
\begin{eqnarray} \label{W1}
(i \eta^{\a\b} X_\a \partial_\b - \kappa) \Psi (x) = 0 \,,
\end{eqnarray}
where $X_\a$ are generalized Dirac matrices. Owing to the fact that each
spinorial representation of $\ol{SL}(4,\RR) \subset \ol{GL}(4,\RR)$
contains an infinite set of Lorentz representations, the Lorentz
\mbox{spinor $\Psi$} will be the infinite sum of spinors $\Psi^{(j)}$.
Each spinor $\Psi^{(j)}$ is chosen in such a way that it describes a
physical spin $j$ particle and/or a resonance on a certain Regge
trajectory. The matrices $X_\a$ contain on its block-diagonal the
$\gamma$-matrices for fermions with spin $1/2, 3/2, 5/2$ etc. The key
ingredient used in this work that accounts for the physically correct 
particle interpretation (e.g.\ proton does not get spin excited by 
boosting) is provided by the deunitarizing automorphisms, a special
feature of the (general) linear groups \cite{Sijacki87}.

In section 7 and 8, we embed the representation used in (\ref{W1}) into
(infinitely many) particularly chosen $\ol{SL}(4,\RR)$ irreducible
representations and replace the spinor $\Psi$ by the manifield
$\manifield$. This yields a still Poincar\'e invariant manifield equation
to which an interaction force can be coupled minimally. The latter must
be gravitational -- or at least gravity-like, as for example the
Chromogravity interaction \cite{NeSi92}, which is seen in an effective
QCD approximation in the IR region and mediated by a di-gluon
chromometric field $G_{\m\n} \sim g_{ab} A_\m{}^a A_\n{}^b \ (a,b = 1, 2,
\dots , 8)$. This is due to the fact that the gauge group of gravity
``effectively'' contains a tensor operator, the shear tensor, which is
able to excite the spin in $\Delta j=2$. In comparison with \cite{Cant}
we make use of {\em non-multiplicity-free} representations of
$\ol{SL}(4,\RR)$ which allow a $\ol{SL}(4,\RR)$ vector operator.

In section 9, we summarize the steps which led to our wave equation.  We
also present a spontaneous symmetry breaking scenario of the (special)
affine group with the physical particle content corresponding to the
Poincar\'e subgroup unitary irreducible representations. Upon breaking
$\ol{SA}(4, \RR)$ down to the Poincar\'e group, we demonstrate how our
equation is connected to a general affine Dirac-type equation.

\newpage

\section{$\ol{SL}(4, \RR)$ vector operator $\widetilde X_\a$} \label{Sec2}

For the construction of a Dirac-type equation, which is to be
invariant under (special) affine transformations, we have two
possibilities to derive the matrix elements of the generalized
Dirac matrices $\widetilde X_\a$.

We can consider the defining commutation relations of a $\ol{SL}(4,
\RR)$ vector operator $\widetilde X_\a$,
\begin{align}
  [\widetilde X_\g, M_{\a\b}]&=i g_{\g\a} \widetilde X_\b - i g_{\g\b}
  \widetilde X_\a, \label{comm1}\\
  [\widetilde X_\g, T_{\a\b}]&=i g_{\g\a} \widetilde X_\b + i g_{\g\b}
  \widetilde X_\a, \label{comm2}
\end{align}
with $g_{\alpha\beta}$ being structure constants of $\ol{SL}(4, \RR)$.
The generators $L_{\alpha\beta}$ of the group $\ol{SL}(4, \RR)$ can be
splitted into the Lorentz generators $M_{\a\b}:=L_{[\alpha\beta]}$ and
the shear generators $T_{\a\b}:=L_{(\alpha\beta)}$. We obtain the
matrix elements of the generalized Dirac matrices $\widetilde X_\a$ by
solving these relations for $\widetilde X_\a$ in the Hilbert space of
a suitable representation of $\ol{SL}(4, \RR)$.

Alternatively, we can embed $\ol{SL}(4, \RR)$ in $\ol{SL}(5, \RR)$. Let
the generators of $\ol{SL}(5,\RR)$ be $L_A{}^B$, $A,B = 0,...,4$.  Then
we define the $\ol{SL}(4, \RR)$ four-vectors $\widetilde X_\a$, and
$\widetilde Y_\a$ by
\begin{align} \label{XequalL4}
\widetilde X_\a:=L_{4\a}, \quad \widetilde Y_\a:=L_{\a 4}, \qquad
\a=0,1,2,3\,.
\end{align}
The operator $\widetilde X_\a$ ($\widetilde Y_\a$) obtained in this way
fulfills the relations (\ref{comm1}) and (\ref{comm2}) by construction.  It is
interesting to point out that the operator $\widetilde G_\a = \frac{1}{2}
(\widetilde X_\a - \widetilde Y_\a )$ satisfies
\begin{align}
[\widetilde G_\a, \widetilde G_\b] = -i M_{\a\b} \,,
\end{align}
thereby generalizing a property of Dirac's $\g$-matrices. Since
$\widetilde X_\a$, $M_{\a\b}$ and $T_{\a\b}$ form a closed algebra, the
application of $\widetilde X_\a$ on the $\ol{SL}(4,\RR)$ states does not
lead out of the $\ol{SL}(4,\RR)$ representation Hilbert space.

In order to obtain an impression about the general structure of the
matrix $\widetilde X_\a$, let us consider the following embedding of
three finite (tensorial) representations of $SL(4,\RR)$ into one of
$SL(5,\RR)$,
\begin{align}
SL(5,\RR) &\supset SL(4,\RR) \nonumber\\
\underbrace{
\stackrel{15}{
\begin{picture}(18,8)
\put(1,-1){\framebox(16,8)}
\put(9,-1){\line(0,1){8}}
\end{picture}} }_{\varphi_{AB}}
&\supset\,
\underbrace{
\stackrel{10}{
\begin{picture}(18,8)
\put(1,-1){\framebox(16,8)}
\put(9,-1){\line(0,1){8}}
\end{picture} } }_{\varphi_{\a\b}}
\oplus
\underbrace{
\stackrel{4}{
\begin{picture}(18,8)
\put(1,-1){\framebox(16,8)}
\put(9,-1){\line(0,1){8}}$\times$
\end{picture}} }_{\varphi_{\a}}
\oplus
\underbrace{
\stackrel{1}{
\begin{picture}(18,8)
\put(1,-1){\framebox(16,8)}
\put(9,-1){\line(0,1){8}}$\times$
\put(-1,0){$\times$}
\end{picture}} }_{\varphi} \,,
\end{align}
where $\square$ is the Young tableau for an irreducible vector representation
of $SL(n,\RR)$, $n=4,5$. The effect of the application of the $SL(4,\RR)$
vector $\widetilde X_\a$ on the fields $\varphi, \varphi_{\a}$ and
$\varphi_{\a\b}$ is
\begin{align} \label{Young}
&\stackrel{\widetilde X_\a}{
\begin{picture}(10,8)
\put(1,-1){\framebox(8,8)}
\end{picture}}
\otimes
\stackrel{\varphi}{
\begin{picture}(18,8)
\put(1,-1){\framebox(16,8)}
\put(9,-1){\line(0,1){8}}$\times$
\put(-1,0){$\times$}
\end{picture} }
\,=
\stackrel{\varphi_\a}{
\begin{picture}(10,8)
\put(1,-1){\framebox(8,8)}
\end{picture}} \nonumber\\
&\stackrel{\widetilde X_\a}{
\begin{picture}(10,8)
\put(1,-1){\framebox(8,8)}
\end{picture}}
\otimes
\stackrel{\varphi_\a}{
\begin{picture}(18,8)
\put(1,-1){\framebox(16,8)}
\put(9,-1){\line(0,1){8}}
{$\times$}
\end{picture} }
\,=
\stackrel{\varphi_{\a\b}}{
\begin{picture}(18,8)
\put(1,-1){\framebox(16,8)}
\put(9,-1){\line(0,1){8}}
\end{picture}} \\
&\stackrel{\widetilde X_\a}{
\begin{picture}(10,8)
\put(1,-1){\framebox(8,8)}
\end{picture}}
\otimes
\stackrel{\varphi_{\a\b}}{
\begin{picture}(18,8)
\put(1,-1){\framebox(16,8)}
\put(9,-1){\line(0,1){8}}
\end{picture} }
\,= 0 \,. \nonumber
\end{align}
Other possible Young tableaux do not appear due to the closure of the
Hilbert space. Gathering these fields in a vector $\varphi_M=(\varphi,
\varphi_\a, \varphi_{\a\b})^{\rm T}$, from (\ref{Young}) we can read
off the structure of $\widetilde X_\a$,
\begin{align} \label{Xvector}
\widetilde X_\a=
\left[
\begin{tabular}{c|c|c}
  $0$  &  & \\
\hline
   \begin{picture}(8,30)
\multiput(0,0)(0,7){4}{x}
\end{picture} & \,\raisebox{1.4ex}{${\bf 0}_4$}  & \\
\hline
     &
\begin{picture}(23,72)
\multiput(0,0)(0,7){10}{xxxx}
\end{picture}
 &  \begin{picture}(72,72)
\multiput(0,0)(0,7){10}{}
\put(30,30){${\bf 0}_{10}$}
\end{picture}
\end{tabular}
\right]\,.
\end{align}
It is interesting to observe that $\widetilde X_\a$ has zero matrices
on the block-diagonal which implies that the mass operator $\k$ in an
affine invariant equation of the type (\ref{W1}) must vanish.

This can be proven for a general finite representation of $SL(4,\RR)$.
Let us consider the action of a vector operator on an arbitrary
irreducible representation $D(g)$ of $SL(4,\RR)$ labeled by
$[\l_1,\l_2,\l_3]$,
\begin{align}
  [\l_1, \l_2, \l_3] \otimes [1,0,0] = &\,[\l_1+1, \l_2, \l_3] \oplus
  [\l_1, \l_2+1, \l_3] \,\oplus \nonumber \\ &\,[\l_1, \l_2, \l_3+1]
  \oplus [\l_1-1, \l_2-1, \l_3-1] \,.
\end{align}
None of the resulting representations agrees with the
representation $D(g)$ nor with the {\em contragradient}
representation $D^{\rm T}(g^{-1})$ given by
\begin{align}
[\l_1,\l_2,\l_3]^{\rm c}=[\l_1,\l_1-\l_3,\l_1-\l_2] \,.
\end{align}
For a general (reducible) representation this implies vanishing
matrices on the block-diagonal of $\widetilde X_\a$ by similar
argumentation as (\ref{Young}) led to the structure (\ref{Xvector}).
Let the representation space be spanned by
$\Phi=(\varphi_1,\varphi_2,...)^{\rm T}$ with $\varphi_i$ irreducible.
Now we consider the Dirac-type equation (\ref{W1}) in the rest frame
$p_\m=(E_{(0)},0,0,0)$ restricted to the subspace spanned by $\varphi_i$,
\begin{align}
  E_{(0)} \bra{\varphi_i} \widetilde X^0 \ket{\varphi_i} =  \bra{\varphi_i}
  \k \ket{\varphi_i} =  m_i \,,
\end{align}
where we assumed the operator $\k$ to be diagonal. So the mass $m_i$ and
therewith $\k$ must vanish since $\bra{\varphi_i} \widetilde X^0
\ket{\varphi_i}=0$.  Therefore, in an affine invariant Dirac-type wave
equation, the mass generation is dynamical, i.e.\ it can only be evoked by an
interaction. This agrees with the fact that the Casimir operator of the
special affine group $\ol{SA}(4,\RR)$ vanishes leaving the masses
unconstrained \cite{Lemk}.  So we believe that our statement also holds for
infinite representations of $\ol{SL}(4, \RR)$.

\section{Prerequisites from the representation theory of the Lorentz
  group \cite{Gelf}} \label{Sec3}

In the following three sections we want to find a Dirac-type equation
for particles with arbitrary half-integral spin. Our main concern will
be the construction of the generalized Dirac matrices $X_\a$. The wave
equation should be invariant with respect to Poincar\'e
transformations.  This implies that $X_\a$ shall be a Lorentz vector
operator satisfying
\begin{align}
 [X_\g, M_{\a\b}]&=i g_{\g\a} X_\b - i g_{\g\b} X_\a \,,
\end{align}
with $M_{\a\b}$ being the Lorentz generators.  We obtain the matrix
elements of the generalized Dirac matrices $X_\a$ by solving these
relations for $X_\a$ in the Hilbert space of a suitable representation
of $\ol{SO}(1,3)$.

\subsubsection*{Determination of $X_\a$ by the method of Gel'fand}
The representations of the Lorentz subgroup $\ol{SO}(1,3)$ can either
be labeled by $\tau=[l_0, l_1]$ or by $D(j_1, j_2)$. These labels are
related by
\begin{align} \label{W50}
l_0=j_1-j_2, \quad l_1=j_1+j_2+1,
\end{align}
with $j_1$ and $j_2$ being the eigenvalues of the Casimir operators of
$SU(2) \times SU(2) \simeq \ol{SO}(1,3)$. The total angular momentum
$l$ is constrained by
\begin{align} \label{W51}
  \vert j_1 - j_2 \vert \leq l \leq j_1+j_2, \quad \textmd{i.e. }\vert
  l_0 \vert \leq l \leq l_1-1.
\end{align}

Two representations $\tau = [l_0,l_1]$ and $\tau' = [l'_0,l'_1]$ are
\emph{coupled} by $X_\a$ when
\begin{align}
[l'_0,l'_1]&=[l_0\pm 1, l_1] \quad\textmd{(type A)}\,,\\
[l'_0,l'_1]&=[l_0, l_1 \pm 1] \quad\textmd{(type B)}.
\end{align}
We depicted them by the {\em interlocking} scheme: $$\tau
\longleftrightarrow \tau'.$$

Assume some irreducible Lorentz representations are given. Gel'fand et
al.\ \cite{Gelf} p.271-277 have solved (\ref{comm1}) for $X_\a$. They
find the matrix elements of $X_0$ to be of the form
\begin{align} \label{X0matrix}
\bigbra{j'_1& j'_2 \\ l'& m' } X_0 \bigket{j_1& j_2 \\ l& m}=
c^{\tau\tau'}_{lm;l'm'}=c^{\tau\tau'}_{l} \delta_{ll'}\delta_{mm'} \, .
\end{align}
For $[l_0', l_1']=[l_0+1,l_1]$ the matrices $c_l^{\tau\tau'}$
$(l=|l_0|,...,l_1-1)$ are given by
\begin{align}
\begin{array}{l}
c_l^{\tau\tau'}=c^{\tau\tau'}\sqrt{(l+l_0+1)(l-l_0)}, \\
c_l^{\tau'\tau}=c^{\tau'\tau}\sqrt{(l+l_0+1)(l-l_0)}, \label{W15}
\end{array}
\end{align}
and for $[l_0', l_1']=[l_0,l_1+1]$ by
\begin{align}
\begin{array}{l}
c_l^{\tau\tau'}=c^{\tau\tau'}\sqrt{(l+l_1+1)(l-l_1)}, \\
c_l^{\tau'\tau}=c^{\tau'\tau}\sqrt{(l+l_1+1)(l-l_1)}, \label{W16}
\end{array}
\end{align}
and $c_l^{\tau\tau'}=c_l^{\tau'\tau}=0$ for non-{interlocking}
representations $\tau$ and $\tau'$.  $c^{\tau\tau'}$ and
$c^{\tau'\tau}$ are arbitrary complex numbers. The matrix elements of
$X_1, X_2$ and $X_3$ can be derived straight-forwardly from $X_0$, see
\cite{Gelf}, p.\ 276f.

\subsubsection*{Requirements on the Lorentz representations}

Which class of irreducible representations are suitable for the
description of fermions? Gel'fand et al.\ \cite{Gelf} impose the
following requirements on the Dirac-type equation (\ref{W1}):

a) It shall be invariant under space reflections.  An irreducible
representation of the \emph{complete} Lorentz group induces a
representation of the \emph{proper} Lorentz group. This representation
is either irreducible (Case I) or it breaks up into two irreducible
pieces (Case II). In the first case we have
$\tau=\dot \tau$, where $\dot \tau=\pm[l_0,-l_1]$ is the
\emph{conjugate} representation of $\tau$.  In the second case, $\tau
\oplus \tau'$, we have $\tau'=\dot \tau$ and the condition $c^{\tau
  \tau'} = c^{\dot \tau \dot \tau'}$ for the parameters in $X_0$.

b) There shall exist a non-degenerate invariant Hermitean form. This
guarantees that Eq.\ (\ref{W1}) can be derived from a Lagrangian.  One
requires that $\tau=\tau^*$ or $\dot \tau=\tau^*$, where
$\tau^*=[l_0,-\bar l_1]$ is the adjoint representation of $\tau$. For
the parameters $c^{\tau \tau'}$ we have the condition $c^{\tau
  \tau'}=\pm \bar c^{\tau'{}^* \tau{}^*}$.

The requirements a) and b) impose constraints on the labels $l_0$ and
$l_1$ of the representations $\tau=[l_0,l_1]$. They are satisfied by
the representations
\begin{align}
[\fract{1}{2}, l_1]  \oplus [ -\fract{1}{2}, l_1],& \quad l_1 \textmd{ real}
\,. \label{W8}
\end{align}

c) The particle shall have positive probability (positive ``charge''),
i.e.\
\begin{align}
\int J_0 \, d^3 x = \int \ol{\Psi} X_0 \Psi\, d^3 x > 0,
\end{align}
and energy of both signs in order to describe particles and
antiparticles. Gel'fand's method guarantees this by requiring $X_0$ to
have eigenvalues $\pm 1$ for states corresponding to the spin of the
particle and vanishing eigenvalues for lower spin components. This
will be demonstrated in the following example.

\section{Determination of \boldmath{$X_\a$} exemplified at a spin 5/2 field}
\label{Sec4}

\begin{figure}
\begin{center}
\begin{picture}(140, 130)
\multiput(51,0)(0,40){3}{\line(1,0){38}}
\multiput(51,2)(0,40){1}{\line(1,0){38}}
\multiput(50,1)(0,40){2}{\line(0,1){38}}
\multiput(90,1)(0,40){2}{\line(0,1){38}}
\multiput(1,0)(4,0){12}{\line(1,0){2}}
\multiput(93,0)(4,0){20}{\line(1,0){2}} \put(140,1){l=1/2}
\multiput(1,40)(4,0){12}{\line(1,0){2}}
\multiput(93,40)(4,0){20}{\line(1,0){2}} \put(140,41){l=3/2}
\multiput(0,80)(4,0){12}{\line(1,0){2}}
\multiput(93,80)(4,0){20}{\line(1,0){2}} \put(140,82){l=5/2}
\put(25,42){$\tau_2$} \put(100,42){$\dot \tau_2$} \put(100,2){$\dot \tau_1$,
  $\dot{\bar{\tau}}_1$}
\put(100,82){$\dot \tau_3$} \put(25,82){$\tau_3$} \put(10,2){$\tau_1$,
  $\bar \tau_1$} \put(45,18){\tiny{2}}\put(92,18){\tiny{0}}
\put(45,58){\tiny{1}} \put(68,4){\tiny{3}} \put(68,42){\tiny{1}}
\put(68,82){\tiny{1}} \put(92,58){\tiny{0}}
\end{picture}
\end{center}
\caption{Interlocking scheme for a spin-5/2 particle.} \label{Pic1}
\end{figure}
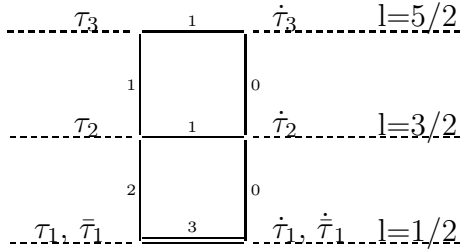
Let us determine the matrix elements of $X_0$ for a fermion with spin
5/2. We follow Gel'fand et al.\ \cite{Gelf} who determined this matrix
for a spin 3/2 particle.  A spin 5/2 particle is described by the four
representations $\tau_1=\bar \tau_1=[\frac{1}{2}, \frac{3}{2}]$,
$\tau_2=[\frac{1}{2}, \frac{5}{2}]$ and $\tau_3=[\frac{1}{2},
\frac{7}{2}]$ and their conjugate representations. $\tau_3$ describes
a composite system of particles with spin 1/2, 3/2 and 5/2. The
representations $\tau_1$, $\bar \tau_1$ and $\tau_2$ are necessary to
eliminate components with spin 1/2 and 3/2 which are introduced by
$\tau_3$. Fig.\ \ref{Pic1} shows the interlocking scheme\footnote{For
  simplicity, arrows indicating interlockings are replaced by lines.}
of these representations. We indicate the presence of two
representations of the same type by a double arrow.

We now want to determine the compartment matrices $c_l^{\tau\tau'}$
which form the Dirac-type matrix $X_0^{(j=5/2)}$, see Eq.\
(\ref{X0matrix}).  From the requirement of parity invariance we obtain:
\begin{align}
c^{\tau_1 \dot \tau_1} = c^{\dot \tau_1 \tau_1}, \quad
c^{\bar \tau_1 \dot{\bar \tau}_1}  = c^{\dot{\bar \tau}_1 \bar \tau_1},\quad
c^{\tau_2 \dot \tau_2} = c^{\dot \tau_2 \tau_2}, \quad
c^{\tau_3 \dot \tau_3} = c^{\dot \tau_3 \tau_3},
\nonumber\\
c^{\tau_1 \dot{\bar\tau}_1} = c^{\dot \tau_1 \bar \tau_1}, \quad
c^{\tau_1 \tau_2} = c^{\dot \tau_1 \dot \tau_2}, \quad
c^{\tau_2 \tau_3} = c^{\dot \tau_2 \dot \tau_3}, \quad
c^{\bar \tau_1 \tau_2} = c^{\dot{\bar \tau}_1 \dot \tau_2},\,
\nonumber\\
c^{\dot{\bar\tau}_1 \tau_1} = c^{\bar \tau_1 \dot \tau_1}, \quad
c^{\tau_2 \tau_1} = c^{\dot \tau_2 \dot \tau_1}, \quad
c^{\tau_3 \tau_2} = c^{\dot \tau_3 \dot \tau_2}, \quad
c^{\tau_2 \bar \tau_1} = c^{\dot \tau_2 \dot{\bar \tau}_1}\,.
\end{align}
From the requirement of the existence of a Hermitean form we get
\begin{align}
c^{\tau_1 \dot \tau_1} = \bar c^{\tau_1 \dot \tau_1}, \quad
c^{\bar \tau_1 \dot{\bar \tau}_1} = \bar c^{\bar \tau_1 \dot{\bar\tau}_1},\quad
c^{\tau_2 \dot \tau_2} = \bar c^{\tau_2 \dot \tau_2}, \quad
c^{\tau_3 \dot \tau_3} = \bar c^{\tau_3 \dot \tau_3}, \nonumber\\
c^{\tau_1 \dot{\bar\tau}_1} = \pm \bar c^{\bar \tau_1 \dot \tau_1}, \quad
c^{\tau_1 \tau_2} = \pm \bar c^{\dot \tau_2 \dot \tau_1}, \quad
c^{\tau_2 \tau_3} = \pm \bar c^{\dot \tau_3 \dot \tau_2}, \quad
c^{\bar \tau_1  \tau_2} =\pm \bar c^{\dot \tau_2 \dot{\bar\tau}_1} .
\end{align}
Using (\ref{X0matrix}) we now compute the compartment matrices
$c_l^{\tau\tau'}$ for $l=1/2, 3/2$, $5/2$ while taking into account the
above relations between the parameters $c^{\tau\tau'}$.  Computer
algebra yields \cite{diplom}:

\begin{minipage}{4cm}
$$
\begin{array}{cc}
\qquad \,\, \tau_3 & \,\, \dot \tau_3
\end{array}
$$
$$
c_{5/2}^{\tau\tau'}= \left[
{\begin{array}{rr}
0 & 3g \\
3g & 0
\end{array}}
 \right],
$$
\end{minipage}
\begin{minipage}{8cm}
$$
\begin{array}{cccc}
\quad \quad \quad \tau_2 & \qquad \dot \tau_2 & \qquad \tau_3
& \qquad \dot \tau_3
\end{array}
$$
$$
c_{3/2}^{\tau\tau'}=%
\left[
{\begin{array}{cccc}
0 & {\displaystyle 2e}  & {\displaystyle \sqrt{\frac{5}{8}}f}  & 0 \\ [2ex]
{\displaystyle 2e}  & 0 & 0 & {\displaystyle \sqrt{\frac{5}{8}}f}  \\ [2ex]
{\displaystyle -\sqrt{\frac{5}{8}}f}  & 0 & 0 & {\displaystyle 2g}  \\ [2ex]
0 & {\displaystyle -\sqrt{\frac{5}{8}}f}  & {\displaystyle 2g}  & 0
\end{array}}
\right] ,
$$
\end{minipage}

\hspace{2cm}
$$
\begin{array}{cccccccc}
\qquad\tau_1 & \dot \tau_1 & \bar \tau_1 & \dot{\bar \tau}_1  &
\tau_2 & \dot \tau_2 & \tau_3 & \dot \tau_3 \\
\end{array}
$$
$$
c_{1/2}^{\tau\tau'}=
\left[
\begin{array}{cccccccc}
0 & a & 0 & b & h & 0 & 0 & 0\\
a & 0 & b & 0 & 0 & h & 0 & 0\\
0 &-b & 0 & c & d & 0 & 0 & 0\\
-b& 0 & c & 0 & 0 & d & 0 & 0\\
h & 0 & d & 0 & 0 & e & f & 0\\
0 & h & 0 & d & e & 0 & 0 & f\\
0 & 0 & 0 & 0 &-f & 0 & 0 & g\\
0 & 0 & 0 & 0 & 0 &-f & g & 0
\end{array}
\right],
$$

where
$$
a=\frac {-1}{3}; b=\frac {1}{24} \,\sqrt{10}; c=\frac {1}{3};
d=\frac {9}{40} \,\sqrt{10}; e=-\frac{1}{3}; f=\frac {4}{15}\, \sqrt{10};
g=\frac{1}{3}; h:=0 .
$$
We excluded particles with spin 1/2 and 3/2 by requiring
additionally that all eigenvalues of the matrices
$c_{1/2}^{\tau\tau'}$ and $c_{3/2}^{\tau\tau'}$ are zero. The
eigenvalues of $c_{5/2}^{\tau\tau'}$ must be $\pm 1$ in order to have
both particles and antiparticles with spin $5/2$.

The compartment matrix $c_{1/2}^{\tau\tau'}$ contains eight
parameters.  Three of them, namely $e, f$ and $g$, are already fixed
by the matrices $c_{3/2}^{\tau\tau'}$ and $c_{5/2}^{\tau\tau'}$. We
can set one parameter equal to zero (here $h=0$) since the requirement
of vanishing eigenvalues fixes only four parameters in the matrix
$c_{1/2}^{\tau\tau'}$. If we had taken just one representation of the
type $\tau_1=[\frac{1}{2}, \frac{3}{2}]$, we would have had only
$5-3=2$ free parameters in $c_{1/2}^{\tau\tau'}$. However, in this
case $3$ parameters will be fixed by the requirement of vanishing
eigenvalues.

\section{Lorentz representation of a fermionic particle} \label{Sec5}

The method of Gel'fand et al.\ can be generalized for all fermions with
spin $j$.  For that we have to use an irreducible Lorentz representation
which contains $j$ as highest spin value. We refer to this as the {\em
main representation}. Thereby we also introduce other spin components
which must be eliminated by a set of auxiliary representations.  The
following theorem helps us to find these representations:

\begin{theorem} \label{TW2}
  The general interlocking scheme for a particle with arbitrary
  half-integral \mbox{spin $j$} reads

\begin{picture}(200, 50)
\multiput(90,11)(50,0){6}{\line(0,1){28}}
\multiput(92,11)(50,0){4}{\line(0,1){28}}
\multiput(94,11)(50,0){1}{\line(0,1){28}}
\multiput(103,5)(50,0){5}{\line(1,0){26}}
\multiput(103,45)(50,0){5}{\line(1,0){26}}
\put(331,2){$\dot \tau_{n+1}$} \put(284,2){$\dot \tau_{n}$}
\put(231,2){$\dot \tau_{n-1}$} \put(181,2){$\dot \tau_{n-2}$}
\put(131,2){$\dot \tau_{n-3}$} \put(81,2){$\dot \tau_{n-4}$}
\put(31,2){$\dot \tau_1$}
\put(54,2){$\cdots$}
\put(331,42){$\tau_{n+1}$} \put(284,42){$\tau_{n}$}
\put(231,42){$\tau_{n-1}$} \put(181,42){$\tau_{n-2}$}
\put(131,42){$\tau_{n-3}$} \put(81,42){$\tau_{n-4}$}
\put(31,42){$\tau_1$}
\put(54,42){$\cdots$}  \put(34,22){$\vdots$}
\end{picture}

where $\tau_i=[\frac{1}{2}, i+\frac{1}{2}]$ and $\dot
\tau_i=[-\frac{1}{2}, i+\frac{1}{2}]$ $(i=1,..., n+1;
n=j-\frac{1}{2})$ are finite irreducible representations of the
Lorentz group. Let us denote the corresponding representation by
$\rho_{j}$. The number $M_i$ of vertical arrows between $\tau_i$ and
$\dot \tau_i$ is the multiplicity with which they occur in $\rho_{j}$.
\end{theorem}

Remarks:
\begin{itemize}
\item[i)] The representations $\rho_{j}$ satisfy the requirements a)
  - c) of Section \ref{Sec3}, i.e.\ Eq.\ (\ref{W1}) is parity invariant,
  derivable from a Lagrangian and describes both particles and
  antiparticles with spin $j$.

\item[ii)] The spin content of the main representation
  $\tau_{n+1}=[\frac{1}{2},j+1]$ is $\frac{1}{2},\frac{3}{2}, ...,j$,
  see Eq.\ (\ref{W51}). The other representations $\tau_1, ...,
  \tau_n$ are needed to eliminate lower spin values such that only a
  particle with spin $j$ remains.

\item[iii)] In Eq.\ (\ref{W1}) we take the field
  $\Psi^{(j)}:=(\psi^{(1)}, ..., \psi^{(i)}, ...,
  \psi^{(n+1)})^T$, where $\psi^{(i)}$ ($i=1,...,n+1$) denotes a
  spinor with (sum over all $j$ values, see ii) )
\begin{align}
\sum\limits_{j=1/2}^{i-1/2} 2\underbrace{(2j+1)}_{\rm m-degeneracy}
=\sum\limits_{l=1}^i 4 j= 2i (i+1)
\end{align}
components. We note that some spinors $\psi^{(i)}$ occur several times in
$\Psi^{(j)}$ according to their multiplicities $M_i$.

\item[iv)] The above interlocking scheme corresponds to the representation
\begin{eqnarray}
\rho_j:=&D(\frac{1}{2}(n+1), \frac{1}{2}n) \oplus
 D(\frac{1}{2}n, \frac{1}{2}(n+1)) \nonumber \\
\oplus &D(\frac{1}{2}n, \frac{1}{2}(n-1)) \oplus
D(\frac{1}{2}(n-1), \frac{1}{2}n) \nonumber \\
 \oplus  &2 [D(\frac{1}{2} (n-1) ,\frac{1}{2}(n-2)) \oplus
D(\frac{1}{2}(n-2), \frac{1}{2} (n-1))]\nonumber\\
 \oplus &2 [D(\frac{1}{2} (n-2) ,\frac{1}{2}(n-3)) \oplus
D(\frac{1}{2}(n-3), \frac{1}{2} (n-2))] \label{representation}\\
 \oplus &2 [D(\frac{1}{2}(n-3) ,\frac{1}{2}(n-4)) \oplus
D(\frac{1}{2}(n-4), \frac{1}{2} (n-3))] \nonumber\\
 \oplus &3 [D(\frac{1}{2}(n-4) ,\frac{1}{2}(n-5)) \oplus
D(\frac{1}{2}(n-5), \frac{1}{2} (n-4))] \nonumber\\
&\vdots \nonumber\\
 \oplus &M_1 [D(\frac{1}{2}, 0) \oplus D(0, \frac{1}{2})]. \nonumber
\end{eqnarray}

\end{itemize}
We will now prove the theorem. The representation (\ref{representation})
shows that the multiplicities $M_i$ do not follow a simple
rule. The proof provides an algorithm for the determination of these
multiplicities.

\subsubsection*{Proof of the theorem}
First, let us assume that a diagram\footnote{For the following we
  rotate it in $90^\circ$ anti-clockwise.} of the type of Theorem
\ref{TW2} is given and we want to verify whether it has the right
number of multiplicities or not. We have to assure that there are
enough parameters to fix in each compartment matrix since then we are
able to set the parameters such that the eigenvalues of the
compartment matrices are all zero and $\pm 1$ for the compartment
matrix with the highest $l$-value, respectively.

This can be achieved in the following way. We write the highest
$l$-value ($=l_1-1$) of each representation $\tau_i$ next to it, see
e.g.\ Fig.\ \ref{Pic1} and Fig.\ \ref{FW1}. Note that the partial
diagram above an $l$-value contains all the information of the number
of parameters of the compartment matrix $c_l^{\tau\tau'}$ with this
$l$-value. It determines the number of free parameters $A_l$ and the
number of parameters $B_l$ which will be fixed by the method of
Gel'fand.

\subsubsection*{Determination of $A_l$:}

Observe that each interlocking gives rise to one parameter
$c^{\tau\tau'}$.  Actually, each interlocking gives rise to two
parameters, $c^{\tau\tau'}$ and $c^{\tau'\tau}$, cf.\ Eqs. (\ref{W15})
and (\ref{W16}).  However, for the class of representations given by
Eq.(\ref{W8}), we have \cite{Gelf} p.320 $\tau=\tau^*$,
$\dot\tau=\dot\tau^*$ and thus $c^{\tau\tau'} = \pm \bar c^{\tau'{}^*
  \tau^*} = \bar c^{\tau'\tau}$ since we have to take into account the
requirement that our Dirac-type equation shall be derivable from a
Lagrangian, cf.\ \cite{Gelf} p.292.  In other words, $c^{\tau\tau'}$
and $c^{\tau'\tau}$ are related. Thus by counting the interlockings of
a partial diagram we obtain the number $A_l$ of parameters in the
compartment matrix $c_l^{\tau\tau'}$.

The number of interlockings $A_l$ can be obtained by counting the arrows
in a diagram. Horizontal arrows are weighted differently than vertical
ones. The following rules can be used to determine these weights.

\textbf{Rule 1 (vertical arrows)}: \emph{Each vertical arrow is weighted by
$n \cdot m$, whereby $n$ and $m$ are the multiplicities of the horizontal
arrows which adjoin it. Vertical arrows between dotted representations are
weighted by zero. }

Example: The following diagram shows a 2-fold and a 3-fold horizontal
arrow.  We weight the vertical arrow by 6 since we get the
parameters $c^{\tau_1 \tau_2}, c^{\tau_1 \tau'_2}, c^{\tau'_1
  \tau_2}$, $c^{\tau'_1 \tau'_2}, c^{\tau''_1 \tau_2}$ and
$c^{\tau''_1 \tau'_2}$, i.e.\ 6 interlockings of type B.
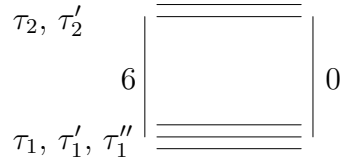
\begin{center}
\setlength{\unitlength}{0.00083333in}
\begingroup\makeatletter\ifx\SetFigFont\undefined%
\gdef\SetFigFont#1#2#3#4#5{%
  \reset@font\fontsize{#1}{#2pt}%
  \fontfamily{#3}\fontseries{#4}\fontshape{#5}%
  \selectfont}%
\fi\endgroup%
{\renewcommand{\dashlinestretch}{30}
\begin{picture}(2042,939)(0,-10)
\path(900,12)(1800,12)
\path(900,87)(1800,87)
\path(900,162)(1800,162)
\path(900,837)(1800,837)
\path(900,912)(1800,912)
\path(1875,837)(1875,87)
\path(825,87)(825,837)
\put(0,762){\makebox(0,0)[lb]{\smash{{{\SetFigFont{12}{14.4}{\rmdefault}{\mddefault}{\updefault}$\tau_2$, $\tau'_2$}}}}}
\put(0,12){\makebox(0,0)[lb]{\smash{{{\SetFigFont{12}{14.4}{\rmdefault}{\mddefault}{\updefault}$\tau_1$, $\tau'_1$, $\tau''_1$}}}}}
\put(675,387){\makebox(0,0)[lb]{\smash{{{\SetFigFont{12}{14.4}{\rmdefault}{\mddefault}{\updefault}6}}}}}
\put(1950,387){\makebox(0,0)[lb]{\smash{{{\SetFigFont{12}{14.4}{\rmdefault}{\mddefault}{\updefault}0}}}}}
\end{picture}
}
\end{center}
Note that, because of parity invariance, we have $c^{\tau \tau'} =
c^{\dot \tau \dot \tau'}$ and therefore {we must not count arrows
between dotted representations}. This is why we weight them by zero.

\textbf{Rule 2 (horizontal arrows)}: \emph{The number of interlockings (of type A)
given by a $n$-fold arrow is }
\begin{align}
\sum\limits_{i=1}^{n} i = \frac{n(n+1)}{2}.
\end{align}
Example: Let us consider a three-fold arrow (represented by three lines).
We simply count the mutual interlockings of the representations $\tau_i$ and
$\dot \tau_i$.
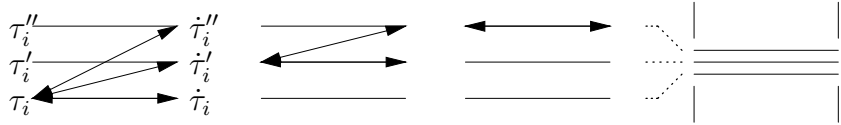
\begin{center}
\setlength{\unitlength}{0.00083333in}
\begingroup\makeatletter\ifx\SetFigFont\undefined%
\gdef\SetFigFont#1#2#3#4#5{%
  \reset@font\fontsize{#1}{#2pt}%
  \fontfamily{#3}\fontseries{#4}\fontshape{#5}%
  \selectfont}%
\fi\endgroup%
{\renewcommand{\dashlinestretch}{30}
\begin{picture}(5187,789)(0,-10)
\path(150,387)(1050,387)
\path(150,162)(1050,162)
\path(150,612)(1050,612)
\path(1575,612)(2475,612)
\path(1575,387)(2475,387)
\path(1575,162)(2475,162)
\path(2850,162)(3750,162)
\path(2850,387)(3750,387)
\path(2850,612)(3750,612)
\blacken\path(259.141,220.209)(150.000,162.000)(273.693,162.000)(259.141,220.209)
\path(150,162)(1050,387)
\blacken\path(940.859,328.791)(1050.000,387.000)(926.307,387.000)(940.859,328.791)
\blacken\path(243.915,242.498)(150.000,162.000)(270.748,188.833)(243.915,242.498)
\path(150,162)(1050,612)
\blacken\path(956.085,531.502)(1050.000,612.000)(929.252,585.167)(956.085,531.502)
\blacken\path(270.000,192.000)(150.000,162.000)(270.000,132.000)(270.000,192.000)
\path(150,162)(1050,162)
\blacken\path(930.000,132.000)(1050.000,162.000)(930.000,192.000)(930.000,132.000)
\drawline(1575,387)(1575,387)
\blacken\path(1695.000,417.000)(1575.000,387.000)(1695.000,357.000)(1695.000,417.000)
\path(1575,387)(2475,387)
\blacken\path(2355.000,357.000)(2475.000,387.000)(2355.000,417.000)(2355.000,357.000)
\blacken\path(1684.141,445.209)(1575.000,387.000)(1698.693,387.000)(1684.141,445.209)
\path(1575,387)(2475,612)
\blacken\path(2365.859,553.791)(2475.000,612.000)(2351.307,612.000)(2365.859,553.791)
\blacken\path(2970.000,642.000)(2850.000,612.000)(2970.000,582.000)(2970.000,642.000)
\path(2850,612)(3750,612)
\blacken\path(3630.000,582.000)(3750.000,612.000)(3630.000,642.000)(3630.000,582.000)
\path(4275,312)(5175,312)
\path(4275,387)(5175,387)
\path(4275,462)(5175,462)
\dottedline{45}(4200,462)(4050,612)(3975,612)
\dottedline{45}(4200,387)(3975,387)
\dottedline{45}(4200,312)(4050,162)(3975,162)
\path(4275,537)(4275,762)
\path(4275,237)(4275,12)
\path(5175,537)(5175,762)
\path(5175,237)(5175,12)
\put(0,537){\makebox(0,0)[lb]{\smash{{{\SetFigFont{12}{14.4}{\rmdefault}{\mddefault}{\updefault}$\tau''_i$}}}}}
\put(0,87){\makebox(0,0)[lb]{\smash{{{\SetFigFont{12}{14.4}{\rmdefault}{\mddefault}{\updefault}$\tau_i$}}}}}
\put(0,312){\makebox(0,0)[lb]{\smash{{{\SetFigFont{12}{14.4}{\rmdefault}{\mddefault}{\updefault}$\tau'_i$}}}}}
\put(1125,537){\makebox(0,0)[lb]{\smash{{{\SetFigFont{12}{14.4}{\rmdefault}{\mddefault}{\updefault}$\dot \tau''_i$}}}}}
\put(1125,312){\makebox(0,0)[lb]{\smash{{{\SetFigFont{12}{14.4}{\rmdefault}{\mddefault}{\updefault}$\dot \tau'_i$}}}}}
\put(1125,87){\makebox(0,0)[lb]{\smash{{{\SetFigFont{12}{14.4}{\rmdefault}{\mddefault}{\updefault}$\dot\tau_i$}}}}}
\end{picture}
}
\end{center}
Here we have altogether six parameters: three parameters $c^{\tau_i
  \dot \tau_i}, c^{\tau_i \dot \tau'_i}, c^{\tau_i \dot \tau''_i}$,
two parameters $c^{\tau'_i \dot \tau'_i}, c^{\tau'_i \dot \tau''_i}$
and one parameter $c^{\tau''_i \dot \tau''_i}$ indicated by arrows
with arrowheads in the above figure. Due to parity invariance
($c^{\tau\tau'}=c^{\dot\tau \dot\tau'}$) there are no further free
parameters. A forth representation $\tau'''_i$ would interlock with
each of the four conjugate representations $\dot \tau_i,\dot \tau'_i,
\dot \tau''_i$, and $\dot \tau'''_i$ and we would obtain four further
interlockings. Clearly, a $n+1$-fold arrow has $n+1$ more
interlockings than the $n$-fold one.

Finally, we apply

\textbf{Rule 3}: \emph{$A_l$ is the sum of the weights of all arrows in a
partial diagram minus the numbers $B_{l'}$ with $l < l' \leq j$. }

When we sum up all weights we get the number of interlockings
and therewith the number of free parameters in the compartment matrix
$c_{l}^{\tau\tau'}$.  We have to subtract $B_{l'}$ ($\forall\, l'>l$)
since these are the number of parameters which have already been fixed
by the compartment matrices $c_{l'}^{\tau\tau'}$ and are not at our
disposal any more.

\subsubsection*{Determination of $B_l$:}

\textbf{Rule 4}: \emph{$B_l$ is the number of horizontal arrows in a partial
diagram --- $n$-fold arrows are counted $n$ times.}

This gives us the number of irreducible representations contributing to
the compartment matrix $c_l^{\tau\tau'}$ and therewith its dimension $n=2
B_l$.  The corresponding characteristic polynomial in $\lambda$ is then
of order $n$ and has the from
\begin{align}
P(\lambda) = \lambda^n+c_{n-2} \lambda^{n-2}+ \cdots + c_2 \lambda^2 + c_0.
\end{align}
Since the eigenvalues of $X_0$ (and therewith those of $c_l^{\tau\tau'}$)
are $\pm \lambda_1, \pm \lambda_2, ...$ \cite{Cors} p.144, the
characteristic polynomial only contains even powers of $\lambda$.  The
constants $c_i$ depend on the parameters of $c_l^{\tau\tau'}$.  In order
to get vanishing eigenvalues, we set the $n/2$ constants $c_i = 0$
($i=0,2,..., n-2$).  These relations fix $n/2=B_l$ parameters.

If for some $l$ the number $A_l$ of parameters which are at our disposal
is less than the number of parameters $B_l$  which will be fixed then the
interlocking scheme fails. In this case we apply

\textbf{Rule 5}: \emph{Assume the multiplicity $M_{l+\frac{1}{2}}$ of the
representation $\tau_{l+\frac{1}{2}}$ is $n$. If $A_l<B_l$, increase
the multiplicity $M_{l+\frac{1}{2}}$ in $1$ by replacing the $n$-fold
by a $n+1$-fold arrow in the diagram and check $A_l$ and $B_l$ again.}

We can always introduce so many representations $\tau_{l+\frac{1}{2}}$
until $A_l \geq B_l$.  Each new representation $\tau_{l+\frac{1}{2}}$
increases $B_l$ in one.  However, $A_l$ increases in
\begin{align}
\frac{(n+1)((n+1)+1)}{2}+ (n+1)\, m - \frac{n(n+1)}{2} - n\,m =
n+1+m > 1 ,
\end{align}
where $n$ is the multiplicity of $\tau_{l+\frac{1}{2}}$ and $m$ that
of $\tau_{l+\frac{1}{2}+1}$, i.e.\ we can always achieve that $A_l
\geq B_l$.

\subsubsection*{Algorithm for obtaining the multiplicities}

Now we are prepared to construct a diagram which has the right
multiplicities.  We start from an ``empty'' diagram, i.e.\ a diagram
with simple arrows everywhere. This corresponds to $j-\frac{1}{2}$
squares.  Using rules 1 and 2, we write the weights next to each arrow
and determine $A_l$ and $B_l$ according to rules 3 and 4.  We begin at
the top horizontal arrow of the diagram: $A_{l=j}=1$ and $B_{l=j}=1$
(o.k. since $A_l \geq B_l$).  Next we evaluate the partial diagram for
$l=j-1$: $A_{l=j-1}=3-1=2$ and $B_{l=j-1}=2$ (o.k.). Then
$A_{l=j-2}=5-2-1=2$ and $B_{l=j-2}=3$ (not o.k.). Therefore, we apply
rule 5, i.e.\ we set $M_{l+\frac{1}{2}=j-\frac{3}{2}}=1 \rightarrow
2$, and get $A_{l=j-2}=8-2-1=5$ and $B_{l=j-2}=4$ (o.k.).  In this way
we go ahead until we reach the bottom arrow.

As in the spin-$\frac{3}{2}$-case \cite{Gelf} p.347f, the energy has both
signs and the charge is positive definite since the compartment matrix with
the highest $l$-value can always be chosen to have eigenvalues $\pm 1$.

Result: By the introduction of enough auxiliary fields it is always
possible to construct a wave equation (\ref{W1}) which fulfills the
wished properties.  \hfill $\square$

In this proof we assumed in Rule 4 that all coefficients $c_i$ of the
characteristic polynomial of a compartment matrix do not vanish ($c_i
\neq 0$) and that they are all different ($c_i \neq c_j$). Of course, it
might be that some $c_i$ are zero in advance or that two or more $c_i$
are equal. Then the relations fix less than $n/2$ parameters. However,
the examples show that this is usually not the case. But this is
difficult to prove. So, strictly speaking, we can only prove that a
scheme works, but not that another one fails. To prove the latter we have
to compute also all characteristic polynomials and check whether there
are $c_i$ which coincide or vanish.

As a final remark we mention that there exists a {\em non-minimal}
solution for the multiplicities. In the appendix we prove that a
representation $\rho_j$ with multiplicities $M_i=n+2-i$ for the
representations $\tau_i$ ($i=1,...,n+1$) can be used for the
description of a particle with spin $j$. So the multiplicities in our
minimal solution given by the above algorithm increase slower than
linearly.
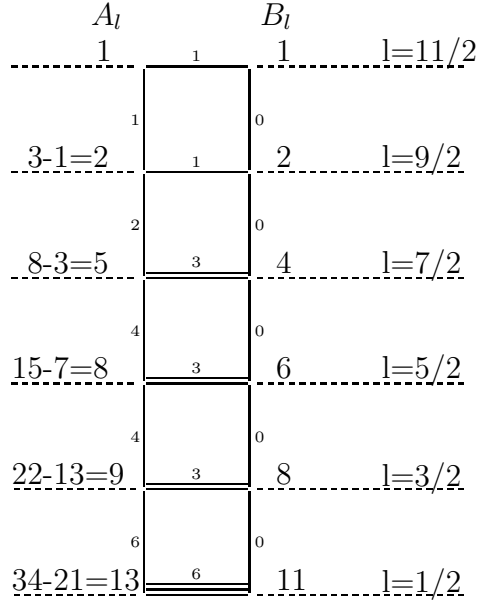
\begin{figure}
\begin{center}
\begin{picture}(140, 250)
\multiput(51,0)(0,40){6}{\line(1,0){38}}
\multiput(51,2)(0,40){4}{\line(1,0){38}}
\multiput(51,4)(0,40){1}{\line(1,0){38}}
\multiput(50,1)(0,40){5}{\line(0,1){38}}
\multiput(90,1)(0,40){5}{\line(0,1){38}}
\multiput(1,0)(4,0){12}{\line(1,0){2}}
\multiput(93,0)(4,0){20}{\line(1,0){2}}
\put(140,1){l=1/2}
\multiput(1,40)(4,0){12}{\line(1,0){2}}
\multiput(93,40)(4,0){20}{\line(1,0){2}}
\put(140,41){l=3/2}
\multiput(0,80)(4,0){12}{\line(1,0){2}}
\multiput(93,80)(4,0){20}{\line(1,0){2}}
\put(140,82){l=5/2}
\multiput(0,120)(4,0){12}{\line(1,0){2}}
\multiput(93,120)(4,0){20}{\line(1,0){2}}
\put(140,122){l=7/2}
\multiput(0,160)(4,0){12}{\line(1,0){2}}
\multiput(93,160)(4,0){20}{\line(1,0){2}}
\put(140,162){l=9/2}
\multiput(0,200)(4,0){12}{\line(1,0){2}}
\multiput(93,200)(4,0){20}{\line(1,0){2}}
\put(140,202){l=11/2}
\put(30,216){$A_l$} \put(94,216){$B_l$}
\put(0,42){22-13=9} \put(100,42){8} \put(100,2){11}
\put(100,82){6} \put(100,122){4} \put(100,162){2} \put(100,202){1}
\put(0,82){15-7=8} \put(6,122){8-3=5} \put(6,162){3-1=2} \put(32,202){1}
\put(0,2){34-21=13}
\put(92,18){\tiny{0}} \put(92,58){\tiny{0}} \put(92,98){\tiny{0}}
\put(92,138){\tiny{0}} \put(92,178){\tiny{0}}
\put(45,18){\tiny{6}} \put(45,58){\tiny{4}} \put(45,98){\tiny{4}}
\put(45,138){\tiny{2}} \put(45,178){\tiny{1}}
\put(68,6){\tiny{6}} \put(68,44){\tiny{3}} \put(68,84){\tiny{3}}
\put(68,124){\tiny{3}} \put(68,162){\tiny{1}} \put(68,202){\tiny{1}}
\end{picture}
\end{center}
\caption{Diagram for a spin-11/2 particle.} \label{FW1}
\end{figure}

\subsubsection*{Comparison with the approach of Singh and Hagen \cite{Sing}}
The aim of our approach is the same as that of Singh-Hagen, though
achieved by a completely different method: We have found a Dirac-type
equation, derivable from a Lagrangian, which replaces the
Rarita-Schwinger scheme of a fermion, see \cite{Sing} Eq.\ (2).
Singh-Hagen found a set of first-order differential equations which
does the same.  Up to spin 9/2 particles both approaches use the same
Lorentz representations, cf.\ the representation (\ref{representation}) with
that in \cite{Sing}.

The interlocking scheme for a spin-11/2-particle is shown in Fig.\
\ref{FW1}.  The number next to each arrow is the number of
interlockings which are induced by it, use rules 1 and\mbox{ 2}.  The above
described method yields three times the representation $\tau_1$ in
contradiction to what Singh-Hagen \cite{Sing} claim.  If we took
$\tau_1$ only twice, as they do, we would obtain $A_{1/2} = 29-21 = 8$
and $B_{1/2}=10$ ($A < B$, not o.k.).  Therefore, we have to introduce
a third $\tau_1$ representation and obtain $A_{1/2} = 34-21 = 13 \geq
B_{1/2} = 11$ (o.k.).

\section{``Reggeization''  } \label{Sec6}
We want to find the Lorentz representation of the resonances on hadronic
Regge trajectories. These resonances can be classified by the group
$\ol{SL}(4,\RR)$ \cite{SiNeD}. When plotted in a Chew-Frautschi diagram,
the Regge trajectories show a linear relation between the square of the
mass $M$ of a resonance and its spin $J$,
\begin{align}
J=\alpha(0)+ \alpha' M^2,
\end{align}
where $\alpha(0)$ sets the low-energy scale, about $1\,GeV$, and
$\alpha'$ is the slope of the trajectories, about $0.9\,(GeV)^{-2}$
(numerical values for the first three flavors).

The extra-ordinary linearity of these trajectories suggests that the
higher spin resonances should rather be described as excitations of
the lowest state of a multiplet than by independent wave equations.
For such a description we define the ``Regge'' representation as the
direct sum of the representations $\rho_j$ given by Theorem \ref{TW2},
\begin{align} \label{reggerep}
  \rho:=\sum_{j=\frac{1}{2}}^{\infty} \!\raisebox{0.8ex}{${}^\oplus$}
  \rho_j \, .
\end{align}
The corresponding infinite-component spinor is $\Psi:=(\Psi^{(1/2)},
\Psi^{(3/2)}, ...)^T$.

The representation $\rho$ describes two exchange-degenerate Regge
trajectories at once: the lowest state of the first one has spin
$\frac{1}{2}$, the other one spin $\frac{3}{2}$. They obey the $\Delta
J=2$ rule, e.g.\ for spinors $\{J\}=\{\frac{1}{2},\frac{5}{2},...\}$
and $\{J\}=\{\frac{3}{2},\frac{7}{2},...\}$. We could also consider
just one Regge trajectory. There is no crucial difference since the
same irreducible Lorentz representations are used.

\begin{figure}
\begin{center}
\setlength{\unitlength}{0.00083333in}
\begingroup\makeatletter\ifx\SetFigFont\undefined%
\gdef\SetFigFont#1#2#3#4#5{%
  \reset@font\fontsize{#1}{#2pt}%
  \fontfamily{#3}\fontseries{#4}\fontshape{#5}%
  \selectfont}%
\fi\endgroup%
{\renewcommand{\dashlinestretch}{30}
\begin{picture}(2988,2859)(0,-10)
\put(1800,1500){\blacken\ellipse{76}{76}}
\put(1800,1500){\ellipse{76}{76}}
\put(1200,900){\blacken\ellipse{76}{76}}
\put(1200,900){\ellipse{76}{76}}
\put(900,600){\blacken\ellipse{76}{76}}
\put(900,600){\ellipse{76}{76}}
\put(1500,1200){\blacken\ellipse{76}{76}}
\put(1500,1200){\ellipse{76}{76}}
\put(2100,1800){\blacken\ellipse{76}{76}}
\put(2100,1800){\ellipse{76}{76}}
\put(900,1163){\ellipse{76}{76}}
\put(1200,1463){\ellipse{76}{76}}
\put(1500,1763){\ellipse{76}{76}}
\put(2400,2100){\blacken\ellipse{76}{76}}
\put(2400,2100){\ellipse{76}{76}}
\put(600,300){\blacken\ellipse{76}{76}}
\put(600,300){\ellipse{76}{76}}
\put(1800,2063){\ellipse{76}{76}}
\put(2100,2363){\ellipse{76}{76}}
\put(600,863){\ellipse{76}{76}}
\put(300,563){\ellipse{76}{76}}
\path(1500,375)(1500,300)
\path(900,375)(900,300)
\path(2100,375)(2100,300)
\path(943.934,888.640)(1050.000,825.000)(986.360,931.066)
\path(1050,825)(825,1050)
\path(975,750)(750,975)
\path(856.066,911.360)(750.000,975.000)(813.640,868.934)
\path(300,900)(375,900)
\path(300,1500)(375,1500)
\path(300,2100)(375,2100)
\path(300,150)(300,2625)
\blacken\path(330.000,2505.000)(300.000,2625.000)(270.000,2505.000)(300.000,2469.000)(330.000,2505.000)
\drawline(300,600)(300,600)
\drawline(1875,2100)(1875,2100)
\path(225,300)(2775,300)
\blacken\path(2655.000,270.000)(2775.000,300.000)(2655.000,330.000)(2619.000,300.000)(2655.000,270.000)
\path(2025,1800)(1800,2025)
\path(1906.066,1961.360)(1800.000,2025.000)(1863.640,1918.934)
\path(675,300)(300,300)(300,675)
\path(1993.934,1938.640)(2100.000,1875.000)(2036.360,1981.066)
\path(2100,1875)(1875,2100)
\dashline{60.000}(300,675)(2250,2625)
\dashline{60.000}(675,300)(2700,2325)
\put(1275,1350){\makebox(0,0)[lb]{\smash{{{\SetFigFont{12}{14.4}{\rmdefault}{\mddefault}{\updefault}$X_\mu$}}}}}
\put(1050,0){\makebox(0,0)[lb]{\smash{{{\SetFigFont{12}{14.4}{\rmdefault}{\mddefault}{\updefault}3/2}}}}}
\put(1650,0){\makebox(0,0)[lb]{\smash{{{\SetFigFont{12}{14.4}{\rmdefault}{\mddefault}{\updefault}5/2}}}}}
\put(2250,0){\makebox(0,0)[lb]{\smash{{{\SetFigFont{12}{14.4}{\rmdefault}{\mddefault}{\updefault}7/2}}}}}
\put(2850,225){\makebox(0,0)[lb]{\smash{{{\SetFigFont{12}{14.4}{\rmdefault}{\mddefault}{\updefault}$j_1$}}}}}
\put(75,300){\makebox(0,0)[lb]{\smash{{{\SetFigFont{12}{14.4}{\rmdefault}{\mddefault}{\updefault}0}}}}}
\put(225,0){\makebox(0,0)[lb]{\smash{{{\SetFigFont{12}{14.4}{\rmdefault}{\mddefault}{\updefault}0}}}}}
\put(450,0){\makebox(0,0)[lb]{\smash{{{\SetFigFont{12}{14.4}{\rmdefault}{\mddefault}{\updefault}1/2}}}}}
\put(0,525){\makebox(0,0)[lb]{\smash{{{\SetFigFont{12}{14.4}{\rmdefault}{\mddefault}{\updefault}1/2}}}}}
\put(0,1125){\makebox(0,0)[lb]{\smash{{{\SetFigFont{12}{14.4}{\rmdefault}{\mddefault}{\updefault}3/2}}}}}
\put(0,1725){\makebox(0,0)[lb]{\smash{{{\SetFigFont{12}{14.4}{\rmdefault}{\mddefault}{\updefault}5/2}}}}}
\put(0,2325){\makebox(0,0)[lb]{\smash{{{\SetFigFont{12}{14.4}{\rmdefault}{\mddefault}{\updefault}7/2}}}}}
\put(1950,2175){\makebox(0,0)[lb]{\smash{{{\SetFigFont{12}{14.4}{\rmdefault}{\mddefault}{\updefault}}}}}}
\put(225,2700){\makebox(0,0)[lb]{\smash{{{\SetFigFont{12}{14.4}{\rmdefault}{\mddefault}{\updefault}$j_2$}}}}}
\put(450,975){\makebox(0,0)[lb]{\smash{{{\SetFigFont{12}{14.4}{\rmdefault}{\mddefault}{\updefault}$\tau_2$}}}}}
\put(675,1200){\makebox(0,0)[lb]{\smash{{{\SetFigFont{12}{14.4}{\rmdefault}{\mddefault}{\updefault}$\tau_3$}}}}}
\put(975,1500){\makebox(0,0)[lb]{\smash{{{\SetFigFont{12}{14.4}{\rmdefault}{\mddefault}{\updefault}$\tau_4$}}}}}
\put(1275,1800){\makebox(0,0)[lb]{\smash{{{\SetFigFont{12}{14.4}{\rmdefault}{\mddefault}{\updefault}$\tau_5$}}}}}
\put(1650,2175){\makebox(0,0)[lb]{\smash{{{\SetFigFont{12}{14.4}{\rmdefault}{\mddefault}{\updefault}.}}}}}
\put(1725,2250){\makebox(0,0)[lb]{\smash{{{\SetFigFont{12}{14.4}{\rmdefault}{\mddefault}{\updefault}.}}}}}
\put(1800,2325){\makebox(0,0)[lb]{\smash{{{\SetFigFont{12}{14.4}{\rmdefault}{\mddefault}{\updefault}.}}}}}
\put(150,675){\makebox(0,0)[lb]{\smash{{{\SetFigFont{12}{14.4}{\rmdefault}{\mddefault}{\updefault}$\tau_1$}}}}}
\put(675,150){\makebox(0,0)[lb]{\smash{{{\SetFigFont{12}{14.4}{\rmdefault}{\mddefault}{\updefault}$\tau_1$}}}}}
\put(975,450){\makebox(0,0)[lb]{\smash{{{\SetFigFont{12}{14.4}{\rmdefault}{\mddefault}{\updefault}$\tau_2$}}}}}
\put(1275,750){\makebox(0,0)[lb]{\smash{{{\SetFigFont{12}{14.4}{\rmdefault}{\mddefault}{\updefault}$\tau_3$}}}}}
\put(1575,1050){\makebox(0,0)[lb]{\smash{{{\SetFigFont{12}{14.4}{\rmdefault}{\mddefault}{\updefault}$\tau_4$}}}}}
\put(1875,1350){\makebox(0,0)[lb]{\smash{{{\SetFigFont{12}{14.4}{\rmdefault}{\mddefault}{\updefault}$\tau_5$}}}}}
\put(2175,1650){\makebox(0,0)[lb]{\smash{{{\SetFigFont{12}{14.4}{\rmdefault}{\mddefault}{\updefault}.}}}}}
\put(2250,1725){\makebox(0,0)[lb]{\smash{{{\SetFigFont{12}{14.4}{\rmdefault}{\mddefault}{\updefault}.}}}}}
\put(2325,1800){\makebox(0,0)[lb]{\smash{{{\SetFigFont{12}{14.4}{\rmdefault}{\mddefault}{\updefault}.}}}}}
\end{picture}
}  \caption{($j_1$, $j_2$)-content of the Regge
representation.} \label{WF1}
\end{center}
\end{figure}
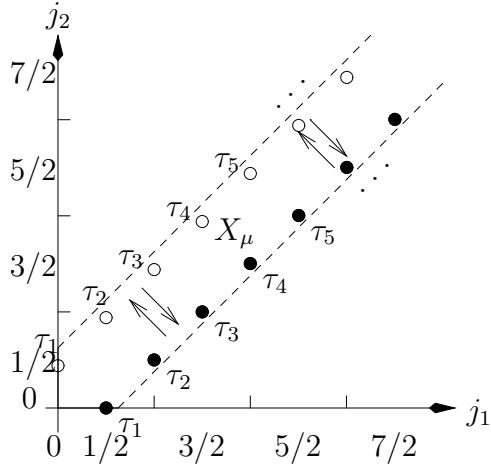
The irreducible representations in (\ref{representation}) are depicted
in Fig.\ \ref{WF1}. All of them lie within the \emph{zone of
  non-trivial action} of $X_\a$.  For example, the representation
$D(\frac{1}{2}, 0) \oplus D(0, \frac{1}{2})$ is indicated by a filled
and an open circle at $(j_1, j_2)= (\frac{1}{2}, 0)$ and $(0,
\frac{1}{2})$.

We can now apply the method of Gel'fand in order to determine the
matrices $X_0^{(j)}$ for each particle with spin $j$. $X_0^{(1/2)}$ is
equal to $\g_0$ used in the conventional Dirac equation. The $X_0$
matrix (and therewith $X_a$ $(a=1,2,3)$) corresponding to the Regge
representation $\rho$ is of the blockdiagonal form
\begin{eqnarray} \label{W11}
X_0=\left[
\begin{array}{ccccc}
X_0^{(1/2)} &           &              \\
          & X_0^{(3/2)} &              \\
          &             & X_0^{(5/2)} &              \\
          &             &             & \ddots  &
\end{array}
\right] \,.
\end{eqnarray}
Thus (\ref{W1}) becomes an infinite set of decoupled equations
describing free Regge resonances.

We now couple the representations $\rho_j$ in order to introduce spin
excitations of the resonances. Physically such excitations of the spin
can only be induced by an interaction force since the spin value does not
change as long as the particle only undergoes Lorentz transformations.
Two neighboring resonances on the Regge trajectories differ in their spin
value in 2. We need an operator which interlocks the representations
$\tau = [\frac{1}{2},l_1]$ and $\tau' = [\frac{1}{2},l_1+2]$. It turns
out that this can be done by the shear operators of the group $\ol{SL}(4,
\RR)$. Since the latter can only act on $\ol{SL}(4, \RR)$ manifields
$\manifield$, we have to embed the Regge representation into a
representation of this group.

\section{(Non-)multiplicity-free representations of\\
  \boldmath{$\ol{SL}(4, \RR)$}}

\subsection{Multiplicity-free representations}
Can we embed the Regge representation into a multiplicity-free ($k_i=0,
\,i=1,2$) representation of $\ol{SL}(4,\RR)$? The multiplicity-free
representations are well known. They have been classified in \cite{Sija85}.
We will not repeat this here, but we strongly recommend to study them
before going ahead.

According to Harish-Chandra\footnote{For a summary of the
representation theory of noncompact groups developed by Harish-Chandra
see also \cite{Sija75} \mbox{Sec.\ 3}.} \cite{Hari} the
representations $U(g)$, $g \in G$ of a noncompact group $G$ can be
defined in a homogeneous Hilbert space $H=\{f(k) \vert k \in K \}$
over the maximal compact subgroup $K \subset G$.  Then $U(g)$ is a
continuous mapping from $G$ into the set of linear transformations on
$H$ given by
\begin{align}
U(g) f(k) = \exp[\a(h(k,g))] f(k \cdot g),
\end{align}
where $g \in G$, $k \in K$, $e^h \in A$ and $A$ is the Abelian
subgroup.  The maximal compact subgroup of $\ol{SL}(4,\RR)$ is
$\ol{SO}(4) \simeq SU(2) \times SU(2)$. After the application of the
\emph{deunitarizing automorphism} $\mathcal{A}$ \cite{Sija85}, the
eigenvalues of its Casimir operators, $j_1$ and $j_2$, can be
identified with those of the Lorentz group since $\ol{SO}(4)_{\cal A}
\simeq \ol{SO}(1,3)$. Each representation of $\ol{SL}(4,\RR)$ contains
Lorentz submultiplets ($j_1$, $j_2$). All these submultiplets are
called the ($j_1$, $j_2$)-content of a $\ol{SL}(4,\RR)$
representation.
\begin{figure}
\begin{center}
\input{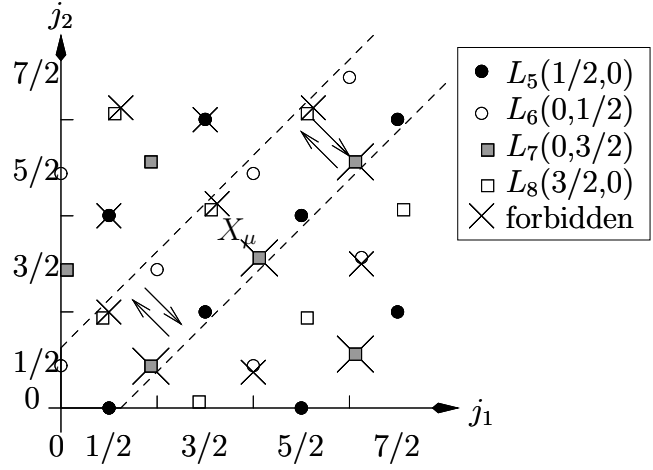} \caption{($j_1$, $j_2$)-content of some multiplicity-free
representations of $\ol{SL}(4,\RR)$.} \label{WF2}
\end{center}
\end{figure}

The Lorentz ($j_1$, $j_2$) submultiplets can be excited by means of the shear
operator $Z_{\a\b}$ $(\a,\b = 0, \pm 1)$, which is in its turn a $(1,1)$
irreducible tensor operator of the Lorentz group. From its matrix
representation in the general case, see (\ref{W20}) below, we deduce that its
action can be visualized by a `Union Jack', for details see \cite{MAG} Ch.\ 
4.5. In Fig.\ \ref{WF3} this is demonstrated for the point $(7/2,1)$.  Due to
the properties of the 3-j-symbols in the multiplicity-free case, we just have
`$\times$'-like transitions between Lorentz submultiplets such that the
lattice is divided into eight sublattices \cite{Sija85} Fig.\ 1. Four of them,
$L_5, L_6, L_7$ and $L_8$, could be relevant for the embedding of the Regge
representation. They are drawn in Fig.\ \ref{WF2}.  However, not all of their
Lorentz submultiplets belong to an \emph{invariant} lattice, i.e.\ to a
multiplicity-free representation of $\ol{SL}(4,\RR)$.  We crossed them out in
Fig.\ \ref{WF2}. By comparison with \mbox{Fig.\ \ref{WF1}} we see that the
irreducible Lorentz representations $D(n,n-1/2) \oplus D(n-1/2,n)$
($n=1,2,...$) of the Regge representation cannot be embedded into any of the
multiplicity-free representations of $\ol{SL}(4,\RR)$. Only those with
$n=\frac{1}{2}, \frac{3}{2},...$ are contained in the lattices $L_5$ and $L_6$
and could be embedded into the $\ol{SL}(4,\RR)$ representation $D^{\rm
  disc}(\frac{1}{2},0)_{\cal A} \oplus D^{\rm disc}(0,\frac{1}{2})_{\cal A} $.

Moreover, we face another problem with {\em multiplicity-free}
representations of $\ol{SL}(4, \RR)$. It is shown in \cite{Cant} App.\ A
that no multiplicity-free representation (except for the sum of ladder
representations which are of no use here) admits an $\ol{SL}(4, \RR)$
vector, i.e. $(\frac{1}{2},\frac{1}{2})$, operator $\widetilde X_\a$.

Indeed, for finite (tensorial) representations this can easily be seen by
using Young tableaux. The tensor product of $\widetilde X_\a$, represented by
$\square$, and a multiplicity-free (all are ladder type) representation
results in the sum of a multiplicity-free and a non-multiplicity-free
representation,
\begin{align}
\underbrace{
\begin{picture}(58,12)
\put(1,3){\framebox(56,8)}
\put(9,3){\line(0,1){8}}
\put(17,3){\line(0,1){8}}
\put(41,3){\line(0,1){8}}
\put(49,3){\line(0,1){8}}
\put(22,4){$\cdots$}
\end{picture}}_{\rm multiplicity-free}
\raisebox{0.8ex}{$\otimes$}\,\,
\begin{picture}(10,12)
\put(1,3){\framebox(8,8)}
\end{picture}
\,\,\raisebox{0.8ex}{$=$}\,\,
\underbrace{
\begin{picture}(66,12)
\put(1,3){\framebox(64,8)}
\put(9,3){\line(0,1){8}}
\put(17,3){\line(0,1){8}}
\put(41,3){\line(0,1){8}}
\put(49,3){\line(0,1){8}}
\put(57,3){\line(0,1){8}}
\put(22,4){$\cdots$}
\end{picture}}_{\rm multiplicity-free}
\,\,\,\,\raisebox{0.8ex}{$\oplus$}
\underbrace{
\begin{picture}(58,16)
\put(1,7){\framebox(56,8)}
\put(9,7){\line(0,1){8}}
\put(17,7){\line(0,1){8}}
\put(41,7){\line(0,1){8}}
\put(49,7){\line(0,1){8}}
\put(22,8){$\cdots$}
\put(1,-1){\framebox(8,8)}
\end{picture}}_{\rm non-multiplicity-free}
\end{align}
Consequently, the application of a $\ol{SL}(4, \RR)$ vector operator
$\widetilde X_\a$ naturally leads to non-multiplicity-free
representations. In the case of spinorial (infinite-dimensional)
representations, we point out two relevant facts: (i) these
representations are not of the ladder type, and (ii) the tensor product
of the vector representation ($\widetilde X_\a$) and a multiplicity-free
spinorial irreducible representation {\em does not} contain any
representation of the latter type. Thus, it is not possible to restrict
on multiplicity-free representations alone.

\subsection{Non-multiplicity-free representations}

Some results for the general case can be found in \cite{Sija89,
  Sija98}. Here the representations are non-multiplicity-free, i.e.\
the label $k_i \neq 0$ ($i=1,2$). The generators of $\ol{SL}(4,\RR)$,
the Lorentz and shear generators, $M_{\a\b}$ and $T_{\a\b}$, can be
replaced by the spherical tensors $J^{(1)}_\a$, $J^{(2)}_\a$, and
$Z_{\a\b}$ $(\a,\b=0,\pm1)$ \cite{Sija85}. The matrix elements of the
$SU(2)$ generators $J^{(1)}_\a$ and $J^{(2)}_\a$ are well known from
angular momentum theory.  The matrix elements of the shear generators
$Z_{\a\b}$ $(\a,\b=0,\pm1)$ read \cite{Sija89}
\begin{align} \label{W20}
\bigbra{j'_1\!&\!j'_2 \\ k'_1 m'_1\! & \!k'_2 m'_2}
        Z_{\a\b} &
\bigket{j_1\!&\!j_2 \\k_1 m_1\!& \!k_2 m_2} =
(-1)^{j'_1-m'_1} \begin{pmatrix} j'_1 & 1 & j_1\\
                                -m'_1 & \a & m_1
                 \end{pmatrix}
                 \times  \\ &\times
(-1)^{j'_2-m'_2} \begin{pmatrix} j'_2 & 1 & j_2\\
                                -m'_2 & \b & m_2
                 \end{pmatrix}\nonumber
                 \bigbra{j_1'&j'_2\\ k'_1\!&\!k'_2} \! \left\vert
                   \begin{matrix} \mbox{}\\\mbox{}
   \end{matrix} Z \right\vert \!
\bigket{j_1&j_2\\ k_1 \!& \!k_2}
\end{align}
with the reduced matrix element
\begin{align}
&\bigbra{j_1'&j'_2\\ k'_1&k'_2} \! \left\vert \begin{matrix} \mbox{}\\\mbox{}
   \end{matrix} Z \right\vert \!
\bigket{j_1&j_2\\ k_1&k_2} = \nonumber\\
 & (-)^{j'_1-k'_1} (-)^{j'_2-k'_2} \fract{i}{2}
   \sqrt{(2j'_1+1)(2j'_2+1))(2j_1+1)(2j_2+1)} \times \nonumber\\
 & \times \left\{\begin{matrix} \mbox{} \\ \mbox{} \end{matrix}
 [e + 4 - j'_1(j'_1+1)+j_1(j_1+1)-j'_2(j'_2+1)+j_2(j_2+1)]
    \right.  \nonumber\\
 &\times \begin{pmatrix} j'_1 & 1 & j_1\\
                          -k'_1 & 0 & k_1
          \end{pmatrix}
          \begin{pmatrix} j'_2 & 1 & j_2\\
                          -k'_2 & 0 & k_2
          \end{pmatrix} \nonumber
\end{align}
\begin{align}
 &-(c+k_1-k2) \begin{pmatrix} j'_1 & 1 & j_1\\
                             -k'_1 & 1 & k_1
          \end{pmatrix}
          \begin{pmatrix} j'_2 & 1 & j_2\\
                          -k'_2 & -1 & k_2
          \end{pmatrix} \nonumber\\
 &-(c-k_1+k2) \begin{pmatrix} j'_1 & 1 & j_1\\
                             -k'_1 &-1 & k_1
          \end{pmatrix}
          \begin{pmatrix} j'_2 & 1 & j_2\\
                          -k'_2 & 1 & k_2
          \end{pmatrix} \nonumber\\
& +(d+k_1+k2) \begin{pmatrix} j'_1 & 1 & j_1\\
                             -k'_1 & 1 & k_1
          \end{pmatrix}
          \begin{pmatrix} j'_2 & 1 & j_2\\
                          -k'_2 & 1 & k_2
          \end{pmatrix} \nonumber\\
& +(d-k_1-k2)\left. \begin{pmatrix} j'_1 & 1 & j_1\\
                             -k'_1 &-1 & k_1
          \end{pmatrix}
          \begin{pmatrix} j'_2 & 1 & j_2\\
                          -k'_2 &-1 & k_2
          \end{pmatrix} \right\}  \,.
\end{align}
In the Appendix we relate the 15 generators $L_{\a\b} = M_{\a\b} +
T_{\a\b}$ to the spherical tensors $J^{(1)}_\a, J^{(2)}_\a$ and
$Z_{\a\b}$ $(\a,\b=0,\pm 1)$.

\begin{figure}
\hspace{1.5cm}
\input{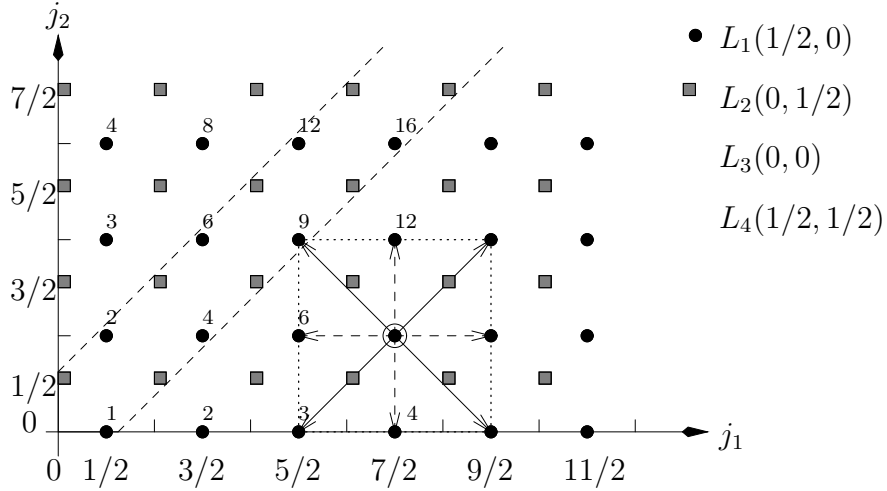} \caption{Four $j_1$-$j_2$-lattices - $L_3$ and $L_4$
are not needed.\hspace{1cm}}
\label{WF3}
\end{figure}

Note some differences to the multiplicity-free case. Since the
operator $Z_{\a\b}$ induces `$\times$'-like \emph{and} `$+$'-like
transitions between Lorentz submultiplets (`Union Jack'), we just have
four sublattices. Two of them, $L_1(\frac{1}{2},0)$ and
$L_2(0,\frac{1}{2})$, which are important for the embedding, are
depicted in \mbox{Fig.\ \ref{WF3}}. Since a state is characterized by
$\ket{j_1 j_2 k_1 k_2}$ and not just by $\ket{j_1 j_2}$ (quantum
numbers $m_1$ and $m_2$ are ignored), we should keep in mind that we
actually deal with a four-dimensional lattice.  Therefore, each dot in
\mbox{Fig.\ \ref{WF3}} can represent more than one Lorentz
submultiplet.  The small-printed number next to each dot is the
multiplicity of the Lorentz subrepresentation $D(j_1,j_2)$.

\subsubsection*{Determination of the multiplicities}
We want to find the multiplicities of the Lorentz submultiplets of
$\ol{SL}(4,\RR)$ representations. As an example, let us determine
those of the lattice $L_1(\frac{1}{2},0)$.  From the properties of the
3-j-symbols in the matrix representation of $Z_{\a\b}$ we know that
$k'_1-k_1= \pm 1$ and $k'_2-k_2= \pm 1$. This allows `$\times$'-like
transitions in the $k_1$-$k_2$-lattice.  It can thus be divided into
eight sublattices in an analogous way as the $j_1$-$j_2$-lattice was
divided in the multiplicity-free case.
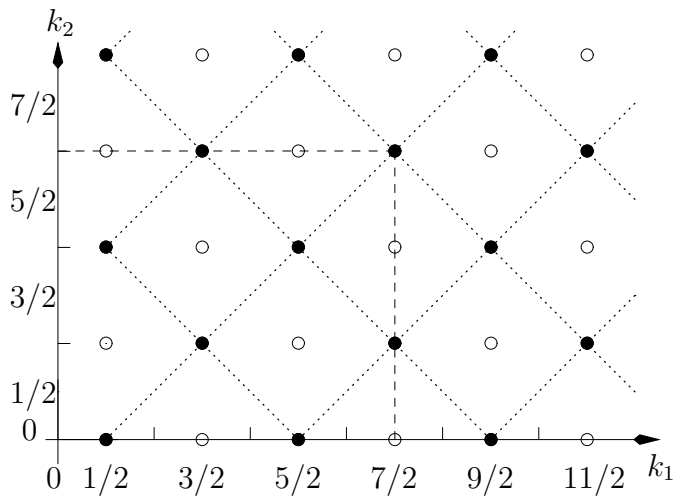
\begin{figure}
\begin{center}
\setlength{\unitlength}{0.00083333in}
\begingroup\makeatletter\ifx\SetFigFont\undefined%
\gdef\SetFigFont#1#2#3#4#5{%
  \reset@font\fontsize{#1}{#2pt}%
  \fontfamily{#3}\fontseries{#4}\fontshape{#5}%
  \selectfont}%
\fi\endgroup%
{\renewcommand{\dashlinestretch}{30}
\begin{picture}(4159,3000)(0,-10)
\put(1800,1500){\blacken\ellipse{76}{76}}
\put(1800,1500){\ellipse{76}{76}}
\put(3000,300){\blacken\ellipse{76}{76}}
\put(3000,300){\ellipse{76}{76}}
\put(3600,900){\blacken\ellipse{76}{76}}
\put(3600,900){\ellipse{76}{76}}
\put(3600,2100){\blacken\ellipse{76}{76}}
\put(3600,2100){\ellipse{76}{76}}
\put(600,300){\blacken\ellipse{76}{76}}
\put(600,300){\ellipse{76}{76}}
\put(600,2700){\blacken\ellipse{76}{76}}
\put(600,2700){\ellipse{76}{76}}
\put(3000,2700){\blacken\ellipse{76}{76}}
\put(3000,2700){\ellipse{76}{76}}
\put(1200,300){\ellipse{76}{76}}
\put(600,900){\ellipse{76}{76}}
\put(2400,2100){\blacken\ellipse{76}{76}}
\put(2400,2100){\ellipse{76}{76}}
\put(1200,2100){\blacken\ellipse{76}{76}}
\put(1200,2100){\ellipse{76}{76}}
\put(600,1500){\blacken\ellipse{76}{76}}
\put(600,1500){\ellipse{76}{76}}
\put(3000,1500){\blacken\ellipse{76}{76}}
\put(3000,1500){\ellipse{76}{76}}
\put(1800,2700){\blacken\ellipse{76}{76}}
\put(1800,2700){\ellipse{76}{76}}
\put(1800,300){\blacken\ellipse{76}{76}}
\put(1800,300){\ellipse{76}{76}}
\put(2400,900){\blacken\ellipse{76}{76}}
\put(2400,900){\ellipse{76}{76}}
\put(1800,900){\ellipse{76}{76}}
\put(3000,900){\ellipse{76}{76}}
\put(3600,1500){\ellipse{76}{76}}
\put(3000,2100){\ellipse{76}{76}}
\put(1800,2100){\ellipse{76}{76}}
\put(2400,1500){\ellipse{76}{76}}
\put(1200,1500){\ellipse{76}{76}}
\put(600,2100){\ellipse{76}{76}}
\put(1200,2700){\ellipse{76}{76}}
\put(2400,2700){\ellipse{76}{76}}
\put(3600,2700){\ellipse{76}{76}}
\put(3600,300){\ellipse{76}{76}}
\put(2400,300){\ellipse{76}{76}}
\put(1200,900){\blacken\ellipse{76}{76}}
\put(1200,900){\ellipse{76}{76}}
\path(1500,375)(1500,300)
\path(2700,375)(2700,300)
\path(300,900)(375,900)
\path(300,1500)(375,1500)
\path(300,2100)(375,2100)
\path(300,150)(300,2775)
\blacken\path(330.000,2655.000)(300.000,2775.000)(270.000,2655.000)(300.000,2619.000)(330.000,2655.000)
\drawline(300,600)(300,600)
\path(900,375)(900,300)
\path(2100,375)(2100,300)
\path(3300,375)(3300,300)
\path(225,300)(4050,300)
\blacken\path(3930.000,270.000)(4050.000,300.000)(3930.000,330.000)(3894.000,300.000)(3930.000,270.000)
\path(675,300)(300,300)(300,675)
\dottedline{45}(600,300)(3150,2850)
\dottedline{45}(3000,300)(600,2700)
\dottedline{45}(600,2700)(750,2850)
\dottedline{45}(3900,1800)(2850,2850)
\dottedline{45}(3000,300)(3900,1200)
\dottedline{45}(3900,600)(1650,2850)
\dottedline{45}(1800,300)(600,1500)
\dottedline{45}(600,1500)(1950,2850)
\dottedline{45}(1800,300)(3900,2400)
\drawline(600,900)(600,900)
\drawline(3900,1200)(3900,1200)
\dashline{60.000}(2400,300)(2400,2100)(300,2100)
\put(75,300){\makebox(0,0)[lb]{\smash{{{\SetFigFont{12}{14.4}{\rmdefault}{\mddefault}{\updefault}0}}}}}
\put(225,0){\makebox(0,0)[lb]{\smash{{{\SetFigFont{12}{14.4}{\rmdefault}{\mddefault}{\updefault}0}}}}}
\put(450,0){\makebox(0,0)[lb]{\smash{{{\SetFigFont{12}{14.4}{\rmdefault}{\mddefault}{\updefault}1/2}}}}}
\put(1050,0){\makebox(0,0)[lb]{\smash{{{\SetFigFont{12}{14.4}{\rmdefault}{\mddefault}{\updefault}3/2}}}}}
\put(1650,0){\makebox(0,0)[lb]{\smash{{{\SetFigFont{12}{14.4}{\rmdefault}{\mddefault}{\updefault}5/2}}}}}
\put(2250,0){\makebox(0,0)[lb]{\smash{{{\SetFigFont{12}{14.4}{\rmdefault}{\mddefault}{\updefault}7/2}}}}}
\put(2850,0){\makebox(0,0)[lb]{\smash{{{\SetFigFont{12}{14.4}{\rmdefault}{\mddefault}{\updefault}9/2}}}}}
\put(3450,0){\makebox(0,0)[lb]{\smash{{{\SetFigFont{12}{14.4}{\rmdefault}{\mddefault}{\updefault}11/2}}}}}
\put(0,525){\makebox(0,0)[lb]{\smash{{{\SetFigFont{12}{14.4}{\rmdefault}{\mddefault}{\updefault}1/2}}}}}
\put(0,1125){\makebox(0,0)[lb]{\smash{{{\SetFigFont{12}{14.4}{\rmdefault}{\mddefault}{\updefault}3/2}}}}}
\put(0,1725){\makebox(0,0)[lb]{\smash{{{\SetFigFont{12}{14.4}{\rmdefault}{\mddefault}{\updefault}5/2}}}}}
\put(0,2325){\makebox(0,0)[lb]{\smash{{{\SetFigFont{12}{14.4}{\rmdefault}{\mddefault}{\updefault}7/2}}}}}
\put(3975,75){\makebox(0,0)[lb]{\smash{{{\SetFigFont{12}{14.4}{\rmdefault}{\mddefault}{\updefault}$k_1$}}}}}
\put(225,2850){\makebox(0,0)[lb]{\smash{{{\SetFigFont{12}{14.4}{\rmdefault}{\mddefault}{\updefault}$k_2$}}}}}
\end{picture}
} \caption{Two of eight $k_1$-$k_2$-lattices.} \label{WF4}
\end{center}
\end{figure}

We now choose two $k_1$-$k_2$-lattices such that they would form the
lattice $L_1(\frac{1}{2},0)$, if the lattices were a $j_1$-$j_2$-lattice
instead of $k_1$-$k_2$-ones. Thus the two relevant $k_1$-$k_2$-lattices
are those shown in Fig.\ \ref{WF4}: one is represented by open circles, the
other one by closed circles.

\begin{figure}
\begin{center}
\setlength{\unitlength}{0.00083333in}
\begingroup\makeatletter\ifx\SetFigFont\undefined%
\gdef\SetFigFont#1#2#3#4#5{%
  \reset@font\fontsize{#1}{#2pt}%
  \fontfamily{#3}\fontseries{#4}\fontshape{#5}%
  \selectfont}%
\fi\endgroup%
{\renewcommand{\dashlinestretch}{30}
\begin{picture}(2838,3009)(0,-10)
\path(300,900)(375,900)
\drawline(300,600)(300,600)
\put(0,525){\makebox(0,0)[lb]{\smash{{{\SetFigFont{12}{14.4}{\rmdefault}{\mddefault}{\updefault}1/2}}}}}
\put(0,1125){\makebox(0,0)[lb]{\smash{{{\SetFigFont{12}{14.4}{\rmdefault}{\mddefault}{\updefault}3/2}}}}}
\path(300,2250)(375,2250)
\drawline(300,1950)(300,1950)
\put(0,1875){\makebox(0,0)[lb]{\smash{{{\SetFigFont{12}{14.4}{\rmdefault}{\mddefault}{\updefault}1/2}}}}}
\put(0,2475){\makebox(0,0)[lb]{\smash{{{\SetFigFont{12}{14.4}{\rmdefault}{\mddefault}{\updefault}3/2}}}}}
\put(600,300){\blacken\ellipse{76}{76}}
\put(600,300){\ellipse{76}{76}}
\put(1800,300){\blacken\ellipse{76}{76}}
\put(1800,300){\ellipse{76}{76}}
\put(2400,2250){\blacken\ellipse{76}{76}}
\put(2400,2250){\ellipse{76}{76}}
\put(1800,2850){\blacken\ellipse{76}{76}}
\put(1800,2850){\ellipse{76}{76}}
\put(1200,2250){\blacken\ellipse{76}{76}}
\put(1200,2250){\ellipse{76}{76}}
\put(1200,900){\blacken\ellipse{76}{76}}
\put(1200,900){\ellipse{76}{76}}
\put(600,900){\blacken\ellipse{76}{76}}
\put(600,900){\ellipse{76}{76}}
\put(1800,900){\blacken\ellipse{76}{76}}
\put(1800,900){\ellipse{76}{76}}
\put(1200,300){\blacken\ellipse{76}{76}}
\put(1200,300){\ellipse{76}{76}}
\put(1800,1500){\blacken\ellipse{76}{76}}
\put(1800,1500){\ellipse{76}{76}}
\put(600,1500){\blacken\ellipse{76}{76}}
\put(600,1500){\ellipse{76}{76}}
\put(1200,1500){\blacken\ellipse{76}{76}}
\put(1200,1500){\ellipse{76}{76}}
\put(1800,2250){\blacken\ellipse{76}{76}}
\put(1800,2250){\ellipse{76}{76}}
\put(1200,2850){\blacken\ellipse{76}{76}}
\put(1200,2850){\ellipse{76}{76}}
\put(2400,2850){\blacken\ellipse{76}{76}}
\put(2400,2850){\ellipse{76}{76}}
\put(2400,900){\blacken\ellipse{76}{76}}
\put(2400,900){\ellipse{76}{76}}
\put(2400,1500){\blacken\ellipse{76}{76}}
\put(2400,1500){\ellipse{76}{76}}
\put(2400,300){\blacken\ellipse{76}{76}}
\put(2400,300){\ellipse{76}{76}}
\path(1500,375)(1500,300)
\path(300,150)(300,1500)
\blacken\path(330.000,1380.000)(300.000,1500.000)(270.000,1380.000)(300.000,1344.000)(330.000,1380.000)
\path(900,375)(900,300)
\path(2100,375)(2100,300)
\path(225,300)(2625,300)
\blacken\path(2505.000,270.000)(2625.000,300.000)(2505.000,330.000)(2469.000,300.000)(2505.000,270.000)
\path(675,300)(300,300)(300,675)
\path(225,1650)(2625,1650)
\blacken\path(2505.000,1620.000)(2625.000,1650.000)(2505.000,1680.000)(2469.000,1650.000)(2505.000,1620.000)
\path(900,1725)(900,1650)
\path(1500,1725)(1500,1650)
\path(2100,1725)(2100,1650)
\path(300,1575)(300,2925)
\blacken\path(330.000,2805.000)(300.000,2925.000)(270.000,2805.000)(300.000,2769.000)(330.000,2805.000)
\drawline(1200,2250)(1200,2250)
\path(1050,2175)(1200,2250)
\path(675,450)(600,300)
\put(225,0){\makebox(0,0)[lb]{\smash{{{\SetFigFont{12}{14.4}{\rmdefault}{\mddefault}{\updefault}0}}}}}
\put(450,0){\makebox(0,0)[lb]{\smash{{{\SetFigFont{12}{14.4}{\rmdefault}{\mddefault}{\updefault}1/2}}}}}
\put(1050,0){\makebox(0,0)[lb]{\smash{{{\SetFigFont{12}{14.4}{\rmdefault}{\mddefault}{\updefault}3/2}}}}}
\put(1650,0){\makebox(0,0)[lb]{\smash{{{\SetFigFont{12}{14.4}{\rmdefault}{\mddefault}{\updefault}5/2}}}}}
\put(2250,0){\makebox(0,0)[lb]{\smash{{{\SetFigFont{12}{14.4}{\rmdefault}{\mddefault}{\updefault}7/2}}}}}
\put(75,2850){\makebox(0,0)[lb]{\smash{{{\SetFigFont{12}{14.4}{\rmdefault}{\mddefault}{\updefault}$j_2$}}}}}
\put(75,1425){\makebox(0,0)[lb]{\smash{{{\SetFigFont{12}{14.4}{\rmdefault}{\mddefault}{\updefault}$j_2$}}}}}
\put(675,525){\makebox(0,0)[lb]{\smash{{{\SetFigFont{12}{14.4}{\rmdefault}{\mddefault}{\updefault}(1/2,0)}}}}}
\put(525,2100){\makebox(0,0)[lb]{\smash{{{\SetFigFont{12}{14.4}{\rmdefault}{\mddefault}{\updefault}(3/2,1)}}}}}
\put(2700,225){\makebox(0,0)[lb]{\smash{{{\SetFigFont{12}{14.4}{\rmdefault}{\mddefault}{\updefault}$j_1$}}}}}
\put(2700,1575){\makebox(0,0)[lb]{\smash{{{\SetFigFont{12}{14.4}{\rmdefault}{\mddefault}{\updefault}$j_1$}}}}}
\put(75,225){\makebox(0,0)[lb]{\smash{{{\SetFigFont{12}{14.4}{\rmdefault}{\mddefault}{\updefault}0}}}}}
\end{picture}
}  \caption{$(j_ 1,j_2)$-sublattices.} \label{WF5}
\end{center}
\end{figure}
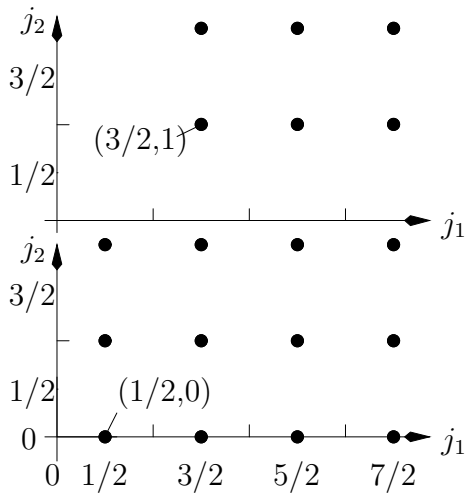

Now, we can ask which $(j_1,j_2)$ submultiplets of
$L_1(\frac{1}{2},0)$ contain a specific pair $(k_1,k_2)$. In other
words, we want to determine the number of states $$\bigket{j_1&j_2\\
  k_1&k_2}$$ for a given pair $(k_1,k_2)$.  Hereto we have to remember
the conditions $j_1 \geq \lvert k_1\rvert$ and $j_2 \geq \lvert
k_2\rvert$.  This means that $(k_1,k_2)$ determines the minimal value
of a sublattice in the $j_1$-$j_2$-lattice in which all $(j_1,j_2)$
submultiplets contain $(k_1,k_2)$. In Fig.\ \ref{WF5} we show two
examples: the $(j_1,j_2)$-sublattices for $(k_1,k_2)=(1/2,0)$ and
$(3/2,1)$.

In order to determine the number of a certain Lorentz submultiplet,
i.e.\ the multiplicity of $(j_1,j_2)$, in principle, we have to determine
the sublattices of the type as in Fig.\ \ref{WF5} for all pairs $(k_1, k_2)$
of the $k_1$-$k_2$-lattices shown in Fig.\ \ref{WF4}. Then we count the number
of sublattices which contain this $(j_1,j_2)$ value. For short, we can
also consider just $(k_1, k_2)=(j_1, j_2)$ in the $k_1$-$k_2$-lattice
and count all the circles which lie inside the rectangle with the
edges $(k_1,k_2)=\{(0,0), (j_1,0), (0,j_2),(j_1,j_2)\}$ since all of them
lead to $(j_1,j_2)$-sublattices which contain this specific $(j_1,j_2)$ value.
In Fig.\ \ref{WF4} this is shown for $(j_1,j_2)=(7/2, 3)$. Its multiplicity
is thus 16. This is the small-printed number next to the component
$(7/2, 3)$ in Fig.\ \ref{WF3}.

We end up with a simple formula for the multiplicity $m$ of a Lorentz
submultiplet $(j_1, j_2)$,
\begin{align}
m = (j_1+a) \times (j_2+b) ,
\end{align}
where $a=b=\frac{1}{2}$ for half-integral and
$a=b=1$ for integral $j_1, j_2$ values.

\section{Embedding of a Regge representation in a \boldmath{$\ol{SL}(4, \RR)$}
  representation} \label{Sec8}

\begin{figure}
\begin{center}
\input{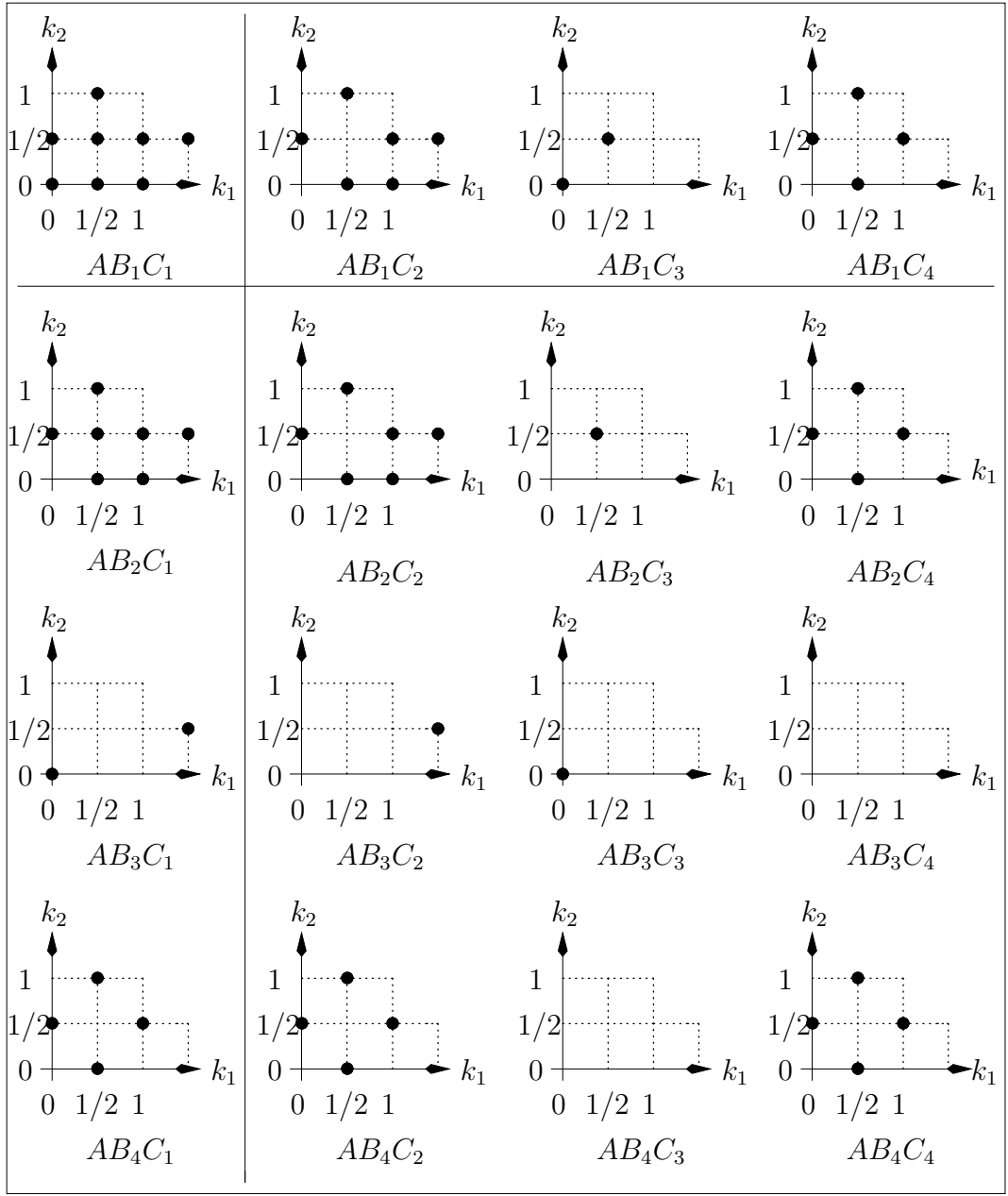}
\caption{The $k_1$-$k_2$-lattice can be divided
into eight sublattices. The 16 diagrams show the possible sublattices
for each series.} \label{WF6}
\end{center}
\end{figure}

The ($j_1$, $j_2$)-content of the Regge representation is shown in
\mbox{Fig.\ \ref{WF1}}. For its embedding we need a series of
$\ol{SL}(4,\RR)$ which contains the $j_1$-$j_2$-lattices
$L_1(\frac{1}{2},0)$ and $L_2(0,\frac{1}{2})$, see \mbox{Fig.\
  \ref{WF3}}. The possible values of the complex representation
labels $c,d,e$ in (\ref{W20}) are \cite{Sija89, Sija98}
\begin{align}
A) &\,e_1=0,\, e_2 \in \RR, \nonumber\\
B_1) &\,d_1=0,\, d_2 \in \RR, \nonumber\\
B_2) &\,d_1=\underline{k}_1 + \underline{k}_2,\, d_2=0; \quad
     \underline{k}_1+\underline{k}_2 = \textstyle\frac{1}{2},1,\frac{3}{2},
     ...,\nonumber\\
B_3) &\,0 < d_1 < 1,\, d_2=0; \quad k_1+k_2=0, \pm2, \pm4, ..., \nonumber\\
B_4) &\,0 < d_1 < \textstyle\frac{1}{2},\, d_2=0; \quad
     k_1+k_2 \equiv\frac{1}{2} (\rm mod\, 2)\,\,\,or\,\,\,\frac{3}{2}(mod\, 2),
\\
C_1) &\, c_1=0,\, c_2 \in \RR, \nonumber\\
C_2) &\, c_1=\underline{k}_1 - \underline{k}_2,\, c_2=0; \quad
     \underline{k}_1-\underline{k}_2 = \textstyle\frac{1}{2},1,\frac{3}{2},
     ...,\nonumber\\
C_3) &\,0 < c_1 < 1,\, c_2=0; \quad k_1-k_2=0, \pm2, \pm4, ..., \nonumber\\
C_4) &\,0 < c_1 < \textstyle\frac{1}{2},\, c_2=0; \quad
     k_1-k_2 \equiv\frac{1}{2} (\rm mod\, 2)\,\,\,or\,\,\,\frac{3}{2}(mod\, 2)
     \,. \nonumber
\end{align}
These are chosen such that the representations are unitary and that
there exists a positive scalar product.  A series of $\ol{SL}(4, \RR)$
is fixed by any combination of $(A)$, $(B_i)$ and $(C_j)$
$(i,j=1,2,3,4)$. For each series one can determine the
$k_1$-$k_2$-sublattices. In principle, there are eight lattices
\begin{align}
L_1&=L(0,0), L_2=L(\fract{1}{2},\fract{1}{2}), L_3=L(0,1)=L(1,0),\nonumber\\
L_4&=L(\fract{1}{2},\fract{3}{2})=L(\fract{3}{2},\fract{1}{2}),
L_5=L(\fract{1}{2},0), L_6=L(0,\fract{1}{2}), \\
L_7&=L(0,\fract{3}{2}), L_8=L(\fract{3}{2},0) \,. \nonumber
\end{align}
In Fig.\ \ref{WF6} only the minimal values $(\underline{k}_1,
\underline{k}_2)$ of these lattices are plotted. All other points of
the $k_1$-$k_2$-lattices can be obtained by performing `$\times$'-like
transitions starting from the minimal values $(\underline{k}_1,
\underline{k}_2)$. For the combination $AB_1C_1$, e.g., we have
neither restrictions on $k_1$ nor on $k_2$. Thus all eight lattices
are allowed, see the first diagram in the upper left corner of Fig.\
\ref{WF6}. While for the series $AB_1C_i$ and $AB_iC_1\, (i=2,3,4)$
there is just one constraint, for the remaining series $k_1$ and $k_2$
have to satisfy two constraints.

Knowing the allowed $k_1$-$k_2$-lattices, we can determine the
$(j_1,j_2)$-content.
Each point $(k_1,k_2)$ denotes all allowed $(j_1,j_2)$, i.e.\ $j_1\geq
\vert k_1 \vert$ and $j_2 \geq \vert k_2 \vert$.

Altogether we find nine series, cf.\ Fig.\ \ref{WF6}, which admit the
$k_1$-$k_2$-lattices $L_5$, $L_6$, $L_7$, and $L_8$.  These lattices
lead to the relevant $j_1$-$j_2$-lattices $L_1(1/2,0)$ and
$L_2(0,1/2)$ of Fig.\ \ref{WF3}.  For example, we could choose the so
called \emph{principal series} - the combination $AB_1C_1$:
\begin{align} \label{pi}
 \pi =
D^{\rm prin}_{\ol{SL}(4,R)}(c_2, d_2, e_2;(\fract{1}{2},0)) \oplus
D^{\rm prin}_{\ol{SL}(4,R)}(c_2, d_2, e_2;(0,\fract{1}{2})).
\end{align}
However, each series, corresponding to one of the combination $AB_iC_j
\,(i,j \neq 3)$ (9 possibilities), can be taken for the embedding.

\section{Dirac-type field equations, minimal coup\-ling of
gravity and symmetry breaking} \label{Ch24}

In this final section we want to review the steps toward affine
generalization of the Dirac equation as well as its coupling to
gravity. Furthermore, we propose a spontaneous symmetry breaking
scenario of the $\ol{SA}(4, \RR)$ gauge symmetry down to the
Poincar\'e one.

We started in flat 4-dimensional Minkowski spacetime. In Sections
\mbox{\ref{Sec3} to \ref{Sec5}} we showed how Gel'fand's method to
derive gamma matrices can be generalized to obtain Dirac-type
equations for fermions with arbitrary spin $j$,
\begin{align}
(i \eta^{\a\b} X^{(j)}_\a \partial_\b  - m^{(j)} ) \, \Psi^{(j)}=0
\,.
\end{align}
The matrix $X^{(j)}_\a$ can be constructed by applying Gel'fand's
method to the representation $\rho_j$ given by Eq.\
(\ref{representation}).

In Section \ref{Sec6} we summed up these representations over all
half-integral spin values, cf.\ Eq.\  (\ref{reggerep}), in order
to describe systems such as two exchange-degenerate Regge
trajectories. Spin excitations of the Regge resonances can then be
introduced by minimal coupling of the Christoffel-type connection
of Chromogravity. This connection is in its turn given in terms of
the chromometric field $G_{\a\b}$, i.e.\ in the anholonomic
notation it reads
\begin{align}
\G^{\{\} \a}_{\b\g} = \frac{1}{2} G^{\a\d} (\partial_{\g}G_{\b\d}
+
\partial_{\b}G_{\g\d} - \partial_{\d}G_{\b\g}) .
\end{align}
The corresponding curved space Dirac-type equation is given by
\begin{align}
(i X^\a e^i{}_\a D_i - \k ) \, \Psi=0 \,, \label{Dtype2}
\end{align}
with the holonomic covariant derivative defined by
\begin{align}
D_i=\partial_i +  \G^{ \{\}}_{i\a\b} L^{\a\b} \,.
\end{align}
This equation is invariant with respect to local Poincar\'e
transformations.

Since the Regge resonances can be classified by the group
$\ol{SL}(4, \RR)$, in Section \ref{Sec8} we embedded the Regge
representation $\rho$ into a suitable representation of
$\ol{SL}(4, \RR)$. Note that the spin content of a genuine world
spinor field is described by the $\ol{SL}(4, \RR)$ representations
as well. Formally, we are now allowed to replace the Lorentz
spinor $\Psi$ in (\ref{Dtype2}) by the manifield $\manifield$
spanning the representation space of a representation of the
series $\pi$ defined in (\ref{pi}). Thus, we obtain a manifield
description that is suitable for either an effective baryonic
field of Regge recurrences or for a world spinor field of affine
gravity.

As argued in Section \ref{Sec2}, in a completely affine wave
equation the mass term vanishes, i.e.\ the equation has to be of
the form
\begin{align}
i \widetilde X^\a e^i{}_\a D_i \manifield =0 \label{Dtype3}
\end{align}
with the $\ol{SL}(4, \RR)$ vector operator $\widetilde X_\a$
defined by (\ref{XequalL4}). In the gravity case, the covariant
derivative $D_i$ now contains a full affine connection which we
take from metric-affine gravity (MAG) \cite{MAG}.

Note that in an equation of the form of Eq.\ (\ref{Dtype2}) we
have not specified the mass term $\kappa$ so far. In order to gain
(\ref{Dtype2}) from (\ref{Dtype3}), we propose, along the lines of
Ref.\ \cite{NeSi88} a symmetry breaking scenario of the $\ol{SA}(4,
\RR)$ which preserves the Poincar\'e symmetry. It is the minimal
spontaneous symmetry breaking scheme in which, besides the
infinite-component $\manifield\!(x)$ field, we introduce an
additional $10$-component second-rank symmetric $\ol{SL}(4,\RR)$
field $\varphi_{\a\b} (x)$. The $\varphi_{\a\b}$ field is the
minimal field that (i) has non-trivial $\ol{SL}(4,\RR)$
transformation properties and (ii) it contains a Lorentz scalar
component, $\varphi^{(0,0)} (x)= \eta^{\a\b} \varphi_{\a\b} (x)$,
thus preserving the Lorentz symmetry in the process of spontaneous
breaking of the $\ol{SL}(4,\RR)$ symmetry. The Lorentz
decomposition of the $\varphi_{\a\b} (x)$ field is $\varphi_{\a\b}
(x) = \varphi^{(0,0)}_{\a\b} (x) + \varphi^{(1,1)}_{\a\b} (x)$,
where $\varphi^{(1,1)}_{\a\b} (x)$ is the traceless $9$-component
field.

We consider the Lagrangian
\begin{align}
{\cal L}={\cal L}_{\rm MAG} \, &+ \conjmanifield i \widetilde X^\a
 e^i{}_\a D_i \!\manifield + \, \frac{1}{2} \eta^{\a\b} e^i{}_\a e^j{}_\b
 (D_i \varphi^{\g\d}) (D_j \varphi_{\g\d}) \nonumber\\ &-
 \m_{\rm M} \conjmanifield  \varphi^{\g\d} \varphi_{\g\d} \manifield
- \frac{\m^2}{2} \varphi^{\g\d} \varphi_{\g\d} - \frac{\l}{4}
(\varphi^{\g\d} \varphi_{\g\d})^2 \,,
\end{align}
which describes manifield $\manifield$, $10$-component field
$\varphi_{\a\b}$, their mutual interaction, as well as their
affine gravity interactions. Here $\varphi_{\a\b}$ interacts with
the manifield with strength $\m_{\rm M}$ and ${\cal L}_{\rm MAG}$
is the most general MAG Lagrangian given by Eq.\ (10) in
\cite{MAG2}. Provided $\m^2 < 0$, one finds a non-trivial vacuum
expectation value determined by
\begin{align}
\l \bra{0} \varphi^{\g\d}\varphi_{\g\d} \ket{0} + \m^2 = 0 .
\end{align}
We perform a suitable $\ol{SL}(4,\RR)$ transformation in the space
of field components, such that $\varphi^{\g\d} \varphi_{\g\d} =
\varphi^{(0,0)\g\d} \varphi^{(0,0)}_{\g\d}$, and obtain the
nontrivial vacuum expectation value for the Lorentz scalar
component, $v \equiv \bra{0} \varphi^{(0,0)} \ket{0} = \sqrt{-\m^2
/ \l}$.

Taking $\varphi_{\a\b} (x) = (v + \chi^{(0,0)}(x))\eta_{\a\b} +
\varphi^{(1,1)}_{\a\b} (x)$, we find a massive scalar field
$\chi^{(0,0)}$, and a set of nine massless Goldstone fields
$\varphi^{(1,1)}_{a\b}$, while the spinorial manifield acquires
mass as well, i.e.
\begin{align}
m(\chi^{(0,0)}) = \sqrt{-2\m^2}\,, \quad m(\varphi^{(1,1)}) = 0\,,
\quad m(\manifield ) = \m_{\rm M} v^2 =: \kappa.
\end{align}

Let us parametrize now $\varphi_{\a\b}$ as follows,
\begin{align}
\varphi_{\a\b} (x) = (v + \chi^{(0,0)}(x)) \eta_{\mu\nu} \,
\exp ({{\textstyle\frac{i}{v}} \chi^{(1,1)}_{\g\d} T^{\g\d}})^\mu{}_{\a}
\exp ({{\textstyle\frac{i}{v}} \chi^{(1,1)}_{\g\d} T^{\g\d}})^\nu{}_{\b} ,
\end{align}
where $T^{\g\d}$ are the shear generators. After the gauge
transformation $U = \exp(-\frac{i}{v} \chi^{(1,1)}_{\g\d}
T^{\g\d})$, the connection fields become (infinitesimally)
\begin{align}
\G'_{i(\a\b)}=\G_{i(\a\b)}-\frac{1}{v} \partial_i
\chi^{(1,1)}_{\a\b} \,,
\end{align}
while the nine Goldstone fields $\chi^{(1,1)}_{\a\b}$ get absorbed by the
symmetric part of the connection $\G_{i(\a\b)}$ which is associated with
nonmetricity. The latter in turn becomes massive, i.e.\ \mbox{$M(\G_{i(\a\b)})
  \neq 0$}.  The antisymmetric part of the connection, which is associated
with spin, remains massless, i.e.\ \mbox{$M(\G_{i[\a\b]}) = 0$}.

We can, furthermore, make use of the nonlinear symmetry
realizations and find explicitly matrix elements of the Lorentz
vector $X^\a_{AB}$ in terms of matrix elements of the
$\ol{SL}(4,\RR)$ vector $\widetilde{X}^\a_{\tilde A
  \tilde B}$, i.e.,
\begin{align}
  &X^\a_{AB} \equiv E^{\tilde C}{}_{A} \widetilde{X}^\a_{\tilde C \tilde D}
  E^{\tilde D}{}_{B}, \quad E^{\tilde A}{}_{B} = \exp (
    {\textstyle\frac{i}{2}} \chi^{(1,1)}_{\a\b} T^{\a\b} )^{\tilde
    A}{}_{B} \,,
  \nonumber\\
  &\Psi_{A} = E^{\tilde A}{}_{A} \manifield_{\tilde A},
\end{align}
where $E^{\tilde A}{}_{B}$, is the nonlinear symmetry realizer. The (tracefree
part of the) MAG-metric tensor $g_{\a\b}$ can be defined from the Goldstone
fields $\chi^{(1,1)}_{\a\b}$ as
\begin{align}
g_{\a\b} := r^\m{}_\a r^\n{}_\b \eta_{\m\n}, \quad r^\m{}_\a:=\exp
({\textstyle\frac{i}{2}} \chi^{(1,1)}_{\a\b} T^{\a\b})^\m{}_\a \,
\end{align}
as suggested by the nonlinear realization of the local affine group
\cite{Tres}.

To summarize, we break spontaneously the $\ol{SL}(4,\RR)$ symmetry
down to the Lorentz symmetry, the fermionic fields acquire nontrivial
mass, and all quantities of an equation of the form given by Eq.\
(\ref{Dtype2}) are explicitly given in terms of the quantities of
\mbox{Eq.\ (\ref{Dtype3})}.

\section*{Acknowledgments}
One of the authors (I.K.) is grateful to F.\ W.\ Hehl (University of
Cologne) for many discussions related to this work.

\section*{Appendix}
\subsection*{Transition from spherical to Cartesian tensors}

It is often useful to relate the Cartesian generators
$L_{\a\b}=M_{\a\b}+T_{\a\b}$ of $\ol{SL}(4,\RR)$ to the spherical
tensors $J^{(1)}_\a, J^{(2)}_\a$ and the double tensor $Z_{\a\b}$
$(\a,\b=0,\pm 1)$.  The inverse of Eq.\ (2.3) in \cite{Sija85} yields
the generators of the maximal compact subgroup $\ol{SO}(4)$,
\begin{align}
&M_{ab}=\ve_{abc} (J^{(1)}_c + J^{(2)}_c), \\
&T_{0a}=J^{(1)}_a - J^{(2)}_a \,.
\end{align}
The relation between the spherical vector $J_{0,\pm}$ and the
Cartesian vector $J_a$ are well-known.

We decompose the double tensor $Z_{\a\b}$ of rank $(1,1)$ with respect to
the rotation group, $\ol{SO}(4) \supset \ol{SO}(3)$,
$
D^{(1)} \times D^{(1)}=D^{(0)} \oplus D^{(1)} \oplus D^{(2)},
$
and obtain the three corresponding tensors
\begin{align}
Z^{(k)}_\g=\sum_{\a,\b} Z_{\a\b}\, (11\a\b|11k\g)\,,
\end{align}
cf.\ Eq.\ (35.2) in \cite{Yuts}, with rank $k=0,1,2$ ($\g=-k,...,+k$)
and the Clebsch-Gordon coefficient $(11\a\b|11k\g)$. The tensors
$Z^{(0)}_\g$, $Z^{(1)}_\g$, and $Z^{(2)}_\g$ have 1, 3, and 5
independent components which we now relate to the Cartesian tensor\mbox{ $Z_{ab}$},
\begin{align}
  &Z_{31}=-\frac{1}{2}(Z^{(2)}_{+1}-Z^{(2)}_{-1}
          +Z^{(1)}_{+1}+Z^{(1)}_{-1})\,,\nonumber\\
  &Z_{13}=-\frac{1}{2}(Z^{(2)}_{+1}-Z^{(2)}_{-1}
          -Z^{(1)}_{+1}-Z^{(1)}_{-1})\,,\nonumber\\
  &Z_{23}=\frac{i}{2}(Z^{(2)}_{+1}+Z^{(2)}_{-1}+Z^{(1)}_{+1}-Z^{(1)}_{-1})\,,
  \nonumber\\
  &Z_{32}=\frac{i}{2}(Z^{(2)}_{+1}+Z^{(2)}_{-1}-Z^{(1)}_{+1}+Z^{(1)}_{-1})\,,
  \nonumber
\end{align}
\begin{align}
 &Z_{12}=-\frac{i}{2}(Z^{(2)}_{+2}-Z^{(2)}_{-2}+\sqrt{2}Z^{(1)}_0) \,,
  \nonumber\\
  &Z_{21}=-\frac{i}{2}(Z^{(2)}_{+2}-Z^{(2)}_{-2}-\sqrt{2}Z^{(1)}_0) \,,
  \nonumber\\
  &Z_{11}=\frac{1}{2}(Z^{(2)}_{+2}+Z^{(2)}_{-2})
         -\frac{1}{\sqrt6} Z^{(2)}_{0}-\frac{1}{\sqrt3}Z^{(0)}_{0}\,,
  \nonumber\\
  &Z_{22}=-\frac{1}{2}(Z^{(2)}_{+2}+Z^{(2)}_{-2})
         -\frac{1}{\sqrt6} Z^{(2)}_{0}-\frac{1}{\sqrt3}Z^{(0)}_{0}\,,
  \nonumber\\
  &Z_{33}=\frac{2}{\sqrt6} Z^{(2)}_{0}-\frac{1}{\sqrt3}Z^{(0)}_{0}\,
 \,.
\end{align}
$Z_{ab}$ is related to the spatial shear tensor $T_{ab}$ and to the
boosts $M_{0c}$ according to
\begin{align}
Z_{ab} = T_{ab} + \ve_{abc} M_{0c}\,.
\end{align}

\subsection*{Non-minimal solution for the multiplicities}

For the multiplicities $M_i=n+2-i$ of the representations $\tau_i$
($i=1,...,n+1$), we have to show that $A_l \geq B_l$ for all
$l=\frac{1}{2},...,j$.

{\bf Proof by induction:} Since $A_j=B_j=1$, $A_l \geq B_l$ for $l=j$.
Now, assume $A_l \geq B_l$. Using Rules 1 to 4, we obtain
\begin{eqnarray}
A_{l-1}&=&A_l+M_i(M_i+1)+\frac{(M_i+1)(M_i+2)}{2}-B_l \nonumber\\
&\geq& (M_i+1)(\frac{3}{2}M_i+1)\\
B_{l-1}&=&B_l+M_i+1
\end{eqnarray}
with $M_i$ being the multiplicity of $\tau_i, i=l+1/2$, and $M_{i-1}=
M_i+1$ the multiplicity of $\tau_{i-1}$.

Now $A_{l-1} \geq B_{l-1}$ follows since
 \begin{eqnarray}
(M_i+1)\frac{3}{2}M_i \geq B_{l}=\sum_{k=i}^{n+1} M_k = \frac{1}{2} M_i(M_i+1)
 \,.
\end{eqnarray}

\end{document}